\documentclass[twocolumn]{aastex701}

\begin{document}
   
\title{The SPACE Program II: No discernible spectral features in the transmission spectrum of the sub-Neptune HD\,191939\,b observed with HST/WFC3}

\author[orcid=0009-0007-9356-8576,gname=Cyril,sname=Gapp]{Cyril Gapp}
\affiliation{Max-Planck-Institut für Astronomie, Heidelberg, Germany}
\affiliation{Department of Physics and Astronomy, Heidelberg University, Heidelberg, Germany}
\email[show]{gapp@mpia.de}

\author[orcid=0000-0002-9147-7925,gname=Lorena,sname='Acuña-Aguirre']{Lorena Acuña-Aguirre}
\affiliation{Max-Planck-Institut für Astronomie, Heidelberg, Germany}
\email{acuna@mpia.de}

\author[orcid=0000-0001-8943-9148,gname=Reza,sname=Ashtari]{Reza Ashtari}
\affiliation{The Johns Hopkins University Applied Physics Laboratory, Laurel, MD, USA}
\email{Reza.Ashtari@jhuapl.edu}

\author[orcid=0000-0002-1830-8260,gname=Mario,sname=Damiano]{Mario Damiano}
\affiliation{Jet Propulsion Laboratory, California Institute of Technology, Pasadena, CA, USA}
\email{mario.damiano@jpl.nasa.gov}

\author[orcid=0000-0001-5442-1300,gname='Thomas M.',sname='Evans-Soma']{Thomas M. Evans-Soma}
\affiliation{School of Science, University of Newcastle, Callaghan, NSW, Australia}
\email{tom.evans-soma@newcastle.edu.au}

\author[orcid=0000-0001-9667-9449,gname='David J.',sname=Wilson]{David J. Wilson}
\affiliation{Laboratory for Atmospheric and Space Physics, University of Colorado Boulder, Boulder, CO, USA}
\email{David.Wilson@lasp.colorado.edu}

\author[orcid=0000-0002-1002-3674,gname=Kevin,sname=France]{Kevin France}
\affiliation{Laboratory for Atmospheric and Space Physics, University of Colorado Boulder, Boulder, CO, USA}
\email{kevin.france@colorado.edu}

\author[orcid=0000-0003-0514-1147,gname=Laura,sname=Kreidberg]{Laura Kreidberg}
\affiliation{Max-Planck-Institut für Astronomie, Heidelberg, Germany}
\email{kreidberg@mpia.de}

\author[gname=Drake,sname=Deming]{Drake Deming}
\affiliation{University of Maryland, College Park, MD, USA}
\email{ldeming@umd.edu}

\author[orcid=0009-0009-9148-2159,gname='Qiushi Chris',sname=Tian]{Qiushi Chris Tian}
\affiliation{Leiden Observatory, Leiden University, Leiden, The Netherlands}
\affiliation{Astronomy Department and Van Vleck Observatory, Wesleyan University, Middletown, CT, USA}
\email{qtian@strw.leidenuniv.nl}

\author[orcid=0000-0003-3786-3486,gname=Seth,sname=Redfield]{Seth Redfield}
\affiliation{Astronomy Department and Van Vleck Observatory, Wesleyan University, Middletown, CT, USA}
\email{sredfield@wesleyan.edu}

\author[gname='Ian J.M.',sname=Crossfield]{Ian J.M. Crossfield}
\affiliation{Department of Physics and Astronomy, University of Kansas, Lawrence, KS, USA}
\email{ianc@ku.edu}

\author[orcid=0000-0001-7714-7551,gname='K. Angelique',sname=Kahle]{K. Angelique Kahle}
\affiliation{Max-Planck-Institut für Astronomie, Heidelberg, Germany}
\affiliation{Department of Physics and Astronomy, Heidelberg University, Heidelberg, Germany}
\email{kahle@mpia.de}

\author[orcid=0000-0002-6939-9211,gname=Tansu,sname=Daylan]{Tansu Daylan}
\affiliation{Department of Physics and McDonnell Center for the Space Sciences, Washington University, St. Louis, MO, USA}
\email{tansu.daylan@gmail.com}

\author[orcid=0000-0003-1907-5910,gname=Kevin,sname=Heng]{Kevin Heng}
\affiliation{Faculty of Physics, Ludwig Maximilian University, Munich, Bavaria, Germany}
\affiliation{Munich Center for Geoastronomy, Ludwig Maximilian University, Munich, Bavaria, Germany}
\affiliation{University College London, Department of Physics \& Astronomy, Gower St, London, UK}
\email{Kevin.Heng@physik.lmu.de}

\author[orcid=0000-0002-8868-7649,gname=Bertram,sname=Bitsch]{Bertram Bitsch}
\affiliation{Department of Physics, University College Cork, College Rd, Cork, Ireland}
\email{bbitsch@ucc.ie}

\author[orcid=0000-0003-2733-8725,gname='James S.',sname=Jenkins]{James S. Jenkins}
\affiliation{Instituto de Estudios Astrofísicos, Facultad de Ingeniería Ciencias, Universidad Diego Portales, Santiago, Chile}
\email{james.jenkins@mail.udp.cl}

\author[orcid=0000-0002-3481-9052,gname='Keivan G.',sname=Stassun]{Keivan G. Stassun}
\affiliation{Department of Physics \& Astronomy, Vanderbilt University, Nashville, TN, USA}
\email{keivan.stassun@vanderbilt.edu}

\author[orcid=0000-0003-1756-4825,gname=Antonio,sname='Garc\'ia Mu\~noz']{Antonio Garc\'ia Mu\~noz}
\affiliation{Universit\'e Paris-Saclay, Universit\'e Paris Cit\'e, CEA, CNRS, AIM, Gif-sur-Yvette, France}
\email{antonio.garciamunoz@cea.fr}

\author[orcid=0000-0001-9355-3752,gname=Ludmila,sname=Carone]{Ludmila Carone}
\affiliation{Space Research Institute, Austrian Academy of Sciences, Graz, Austria}
\email{ludmila.carone@oeaw.ac.at}

\begin{abstract}
The atmospheres of sub-Neptunes provide a window into their internal structure and history, shedding light on the origin of this common, but enigmatic, class of exoplanets. However, the physical and chemical processes that shape sub-Neptunes' transmission spectra, in particular cloud and haze formation, are not well understood. To identify possible correlations between transmission spectra and UV irradiation, the SPACE (Sub-neptune Planetary Atmosphere Characterization Experiment) Program observed an array of sub-Neptunes and their host stars using the Hubble Space Telescope (HST), measuring the planets' transmission spectra between $1.1\,\mu$m and $1.7\,\mu$m with the Wide Field Camera 3 (WFC3) and the stars' UV spectra with the Space Telescope Imaging Spectrograph (STIS). Here, we present the observations of HD\,191939\,b carried out as part of the SPACE Program, which reveal no significant spectral features in the transmission spectrum. The data deliver moderate evidence at significance levels between $2.0\,\sigma$ and $3.2\,\sigma$ against a cloud-free atmosphere with solar metallicity, rendering this scenario unlikely, but still possible. A super-solar metallicity of HD\,191939\,b might be consistent with the known trend of increasing atmospheric metallicity with decreasing planet mass. Both hydrocarbon haze formation and cloud condensation can be efficient at HD\,191939\,b's zero-albedo equilibrium temperature of $(880\pm 20)$\,K, particularly in atmospheres with super-solar metallicity, possibly additionally muting absorption features.
\end{abstract}


\section{Introduction}

Exoplanets smaller than Neptune are by far the most common planets on close orbits around stars in the Galaxy \citep{borucki11,batalha13,fulton17}. Since these planets are such common outcomes of planet formation, understanding their formation and evolution histories offers powerful prospects toward understanding the fundamental principles of planet formation.

\subsection{Insights from the observed planet population}
Two features of the observed occurrence rate of planets with orbital periods shorter than 100 days can be used to explain the origin of planets smaller than 4 Earth radii ($R_\oplus$) on orbits close to their host stars \citep{fulton17}: Firstly, they bifurcate into two distinct populations, commonly referred to as super-Earths with radii between 1 and $1.8\,R_\oplus$ and sub-Neptunes with radii between 1.8 and $4\,R_\oplus$ that are separated by a `radius valley' in between. And secondly, the occurrence rate sharply drops off with increasing planet radius at planet radii larger than $\sim 3R_\oplus$ which is referred to as the `radius cliff'.

Mass loss of planets driven by stellar X-ray and extreme-ultraviolet irradiation \citep{lammer03}, so-called photoevaporation, predicts the radius valley as sub-Neptunes are massive enough to retain primordial hydrogen-helium dominated envelopes, while super-Earths are the left-over cores of planets smaller than sub-Neptunes that lost their envelopes \citep{owen13}. The radius cliff, in turn, indicates that photoevaporation efficiently drives planets around the size of Neptune into the stable sub-Neptune population \citep{owen17}. In an alternative picture, however, both the radius valley and cliff could be the result of planet migration with photoevaporation only playing a minor role in sculpting the observed planet population: Super-Earths are rocky planets that formed inside the snow line and sub-Neptunes accreted large amounts of H$_2$O ice beyond the snow line before migrating inwards \citep{venturini20,luque22,izidoro22,burn24}. The photoevaporation and migration scenarios are not mutually exclusive, though, and thus, the population of close-in exoplanets is possibly the result of both evolution tracks. Sub-Neptunes are the key to quantifying the impact of both scenarios on the whole population, because the migration model, unlike the photoevaporation scenario, requires them to migrate inwards after their formation. Super-Earths, in contrast, can be consistent with a rocky composition and no migration in both cases.

\subsection{Constraining sub-Neptune formation using atmospheric observations}
A promising avenue to constrain a sub-Neptune's formation location in the protoplanetary disk and, with it, its migration history, is to measure the chemical composition of its atmosphere, because the abundances of elements in a planet can reveal the location in the protoplanetary disk it formed at \citep{oberg11,lothringer21}. Multiple challenges, however, complicate the endeavor to connect a sub-Neptune's atmospheric spectrum to its formation location in the protoplanetary disk: Firstly, the envelope masses of sub-Neptunes are unknown, because their masses and radii are consistent with a large variety of possible envelope mass fractions and mean-molecular weights (see, e.g., \citealp{rogers10}). When the mass of the envelope is small compared to the core, its chemical composition can be driven away from its primordial composition inherited from the protoplanetary disk by infalling planetesimals \citep{fortney13}. Secondly, the chemical composition of a sub-Neptune's envelope can be further altered by outgassing from a magma ocean at the core-envelope boundary \citep{misener23,shorttle24,ito25,heng25,werlen25}. This geochemical process can affect some but likely does not affect all sub-Neptunes, because many combinations of sub-Neptune masses and radii disfavor internal magma oceans \citep{breza25}. And thirdly, sub-Neptunes' atmospheric chemistries, in particular the role of aerosols, are not well understood, because a counterpart in the Solar System is missing. Aerosols are challenging for atmospheric characterizations, because they can inhibit measurements of the atmosphere's chemical inventory by obscuring absorption features (see, e.g., \citealp{kreidberg14}).

The impacts of interior-atmosphere interactions and aerosols on sub-Neptunes' transmission spectra likely vary with planetary and stellar parameters, such as the planet's radii ($R_p$) and equilibrium temperatures ($T_{\text{eq}}$) and the host stars' ultraviolet (UV) spectra. Thus, understanding sub-Neptunes' atmospheric spectra as the result of the interplay between these processes and the planet's formation histories requires atmospheric observations on a population level to quantify the effect of each involved process. To that end, past studies have identified correlations between the amplitudes of H$_2$O absorption features in exoplanets' transmission spectra observed with the Wide Field Camera 3 (WFC3) onboard the Hubble Space Telescope (HST) with their $T_{\text{eq}}$. \cite{crossfield17} found muted H$_2$O features at $T_{\text{eq}}\lesssim850$\,K, possibly driven by haze formation and \cite{brande24} identified muted features between $500\,\text{K}\lesssim T_{\text{eq}}\lesssim 700$\,K in line with the earlier study. One of the fundamental limits of population studies, however, is the number of available planetary spectra. For instance, the small sample size of \cite{crossfield17} led to a degeneracy between $T_{\text{eq}}$ and the planets' bulk H$_2$-He mass fractions \citep{lopez14} that could have been the driver of the amplitudes of the observed H$_2$O features instead of $T_{\text{eq}}$. Additionally, the planets analyzed as part of these two past studies spanned radii between 2 and $7.3\,R_\oplus$ to increase the sample size and thus mixed the sub-Neptune population reaching up to $R_p\approx 4\,R_\oplus$ with the likely distinct gas giants. To widen the parameter space of observed sub-Neptunes, the Sub-Neptune Atmospheric Characterization Experiment (SPACE) observed numerous exoplanets with radii between 2 and $4.6\,R_\oplus$ and their host stars using HST (programs GO 17192, GO 17414, PI: Kreidberg), measuring the planets' transmission spectra with WFC3 and the host stars' UV spectra with the Space Telescope Imaging Spectrograph (STIS). \cite{kahle25} presented the SPACE observations of HD\,86226\,c and the SPACE target that is analyzed here is HD\,191939\,b.

\subsection{The HD\,191939 system}
HD\,191939 is a system 53.61\,pc away with a systemic radial velocity of $(-9.52 \pm 0.19)$\,km\,s$^{-1}$ \citep{gaia_dr2}. It hosts six planets orbiting a late G-type central star \citep{badenas_agusti20,lubin22} of which three are transiting sub-Neptunes: HD\,191939\,b, c, and d which have orbital periods of $8.88$, $28.58$ and $38.35$ days, respectively \added{\citep{badenas_agusti20}}. The three non-transiting planets are HD\,191939\,e, g and f which have minimum masses of $0.353\pm 0.013$, $0.0425\pm 0.0063$ and $2.88\pm 0.26$ Jupiter masses and orbital periods of $101.12\pm 0.13$, $284^{+10}_{-8}$ and $2898\pm 152$ days, respectively \citep{lubin22,orellmiquel23,lubin24}.
\begin{figure*}[htbp!]
	\centering
	\includegraphics[width=\hsize]{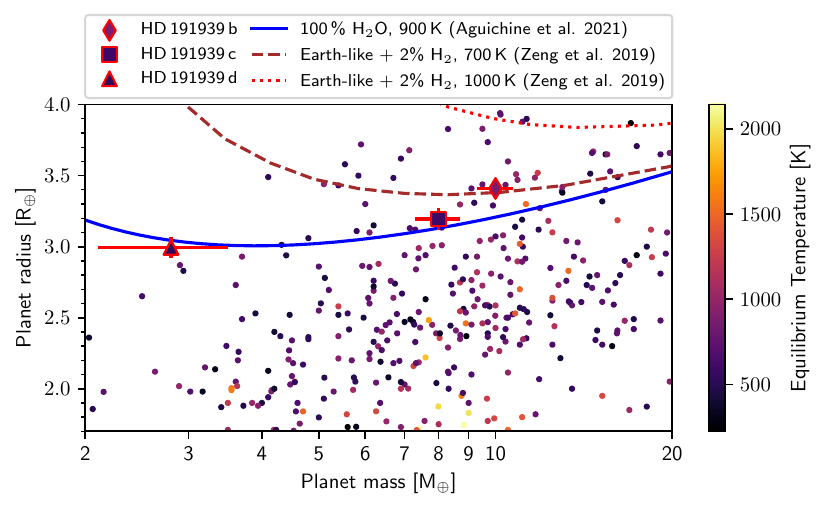}
	\caption{Mass-radius diagram of known exoplanets together with model curves taken from \cite{aguichine21} and \cite{zeng19}. Exoplanet data shown with colored dots were taken from the NASA Exoplanet Archive \citep{exoplanetarchive} on April 2, 2026. The transiting exoplanets in the HD\,191939 system are highlighted with larger markers, error bars and a red outline.}
	\label{fig:population}
\end{figure*}

As a multi-planet system with inner sub-Neptunes and two outer gas giants, HD\,191939 is a system that might yield insight into how protoplanetary disk dynamics and planet-disk interactions can impact planet formation and evolution, particularly for sub-Neptunes. Several models have shown that the inner discs should be enriched in volatiles compared to stellar values due to inward drifting and evaporating pebbles (see, e.g., \citealt{Booth2017, Schneider2021, Mah2023}), which is reflected in inner disc observations (see, e.g., \citealt{Perotti2023, Banzatti2023}) that found large H$_2$O contents. One would assume that planets accreting gas in these inner disc regions would thus be enriched in volatiles (especially H$_2$O), leading to atmospheres with increased mean molecular weights and low carbon-to-oxygen ratios (C/O). However, the HD\,191939 system contains a wide orbit giant planet that could have hindered the inward drift of pebbles (e.g. \citealt{Paardekooper2006, Lambrechts2014, Bitsch2018}) and thus blocked H$_2$O ice rich material from entering the inner system (see, e.g., \citealt{Bitsch2021, Kalyaan2021, Mah2024}). Thus, planets accreting gas in the inner disc regions might only accrete small amounts of volatiles that are dominated by inward moving carbon vapor rich molecules such as CO$_2$, CO and CH$_4$, leading to atmospheres that have a lower mean molecular weight and high C/O (e.g. \citealt{Bitsch2021}). Consequently, determining the composition of inner sub-Neptunes in systems with and without outer giants would be valuable to understand the different formation pathways. Therefore, characterizing the atmospheres of the transiting sub-Neptunes in the HD\,191939 system offers the rare opportunity to test the hypothesis that wide-orbit gas giants might correlate with high C/O in inner sub-Neptunes.

With the shortest orbital period, HD\,191939\,b is the planet with the highest zero-albedo $T_{\text{eq}}$ of $(880\pm20)$\,K in the system \citep{orellmiquel23}. With a mass of $10.00\pm0.70$ Earth masses ($M_\oplus$) and a radius of $(3.410\pm0.075)\,R_\oplus$, HD\,191939\,b's density is too low even for high volatile bulk mass fractions and thus, the planet has to be made of at least some fractions of H$_2$ and He (see Figure \ref{fig:population}). These constraints on the bulk composition increase the prospects for an atmosphere with a low mean molecular weight (see also \citealp{heng25}) and thus large atmospheric absorption features in the planet's transmission spectrum that might be used to measure the chemical composition of its atmosphere.

\section{Observations and data analysis}
To characterize HD\,191939\,b and its host star, the system was observed with HST as part of the SPACE program. The primary transits of HD\,191939\,b on October 17, 2022, November 21, 2022 and July 19, 2023 (UTC, hereafter Visits 1, 2 and 3) were observed with the Wide Field Camera 3 (WFC3) in the infrared (IR) employing its G141 grism to obtain the planet's transmission spectrum between $\lambda\sim 1.1\,\mu$m and $\lambda\sim 1.7\,\mu$m. On December 4, 2022, one observation each of the star without planetary transits was conducted with the G140L, G140M and G230L grisms of the Space Telescope Imaging Spectrograph (STIS) to measure its UV spectrum. UV irradiation is essential for the formation of atmospheric hazes \citep{yung84} and thus, the host star's UV spectrum might be vital for future models of aerosol formation in the planet's atmosphere. Additionally, the optical photometric magnitudes of the host star were monitored between March 2023 and April 2024 using the automated 24-inch telescope at Van Vleck Observatory, Wesleyan University. The monitoring is helpful for detecting stellar variability driven by activity, which is important for the characterization of HD\,191939\,b's atmosphere, since thermal heterogeneity in HD\,191939\,A's photosphere due to, e.g. star spots or faculae, can contaminate the planet's transmission spectrum (see, e.g., \citealp{rackham19}).

\subsection{HST/WFC3 IR observations}
To maximize the fraction of the observing time spent collecting photons, the WFC3 IR time-series observations were conducted using spatial scanning to spread the incoming radiation over more detector pixels in the cross-dispersion direction than without spatial scanning (see, e.g., \citealp{deming13}). All exposures had a duration of 69.617\,s and the telescope was slewed with a scan rate of 0.36"s$^{-1}$ during the exposures, thus spreading the trace over a scan height of 25.062" which corresponds to about 206 pixels on the detector. The scan directions were alternated between forward and reverse to avoid slew times between exposures.

Each of the three transit observations consisted of four HST orbits with 22 exposures per orbit. We reduced and analyzed the observations with three independent reduction pipelines to explore the impact of pipeline-level assumptions on HD\,191939\,b's transmission spectrum.

\subsubsection{\texttt{PACMAN} reduction} \label{subsec:pacman}
\texttt{PACMAN} \citep{pacman} is a data reduction pipeline for reducing and analyzing HST/WFC3 exoplanet observations to generate planetary spectra starting with calibrated intermediate MultiAccum (ima) files provided through the Barbara A. Mikulski Archive for Space Telescopes (MAST). Several HST observations have already been analyzed with \texttt{PACMAN} \citep{bachmann25,kahle25} and parts of the code that were included into the pipeline were previously applied to earlier data analyses (e.g., \citealp{kreidberg14,kreidberg18}). Here, we run the full pipeline starting with pre-processing steps (Stages 00, 01, 02) and a wavelength calibration  (Stage 03), then extracting the white and spectroscopic light curves (Stages 10, 20 and 21) and finally fitting the light curves (Stage 30) to obtain HD\,191939\,b's transmission spectrum.

For the wavelength calibration (Stage 03), we generated a model spectrum of HD\,191939\,A using the 1993 Kurucz Stellar Atmospheres Atlas and adopting an effective temperature $T_{\text{eff}}=5348$\,K, a $\log\,g=4.3$\,cm\,s$^{-2}$ and a metallicity of $[\text{Fe}/\text{H}]=-0.15$\,dex \citep{lubin22}. In Stage 21, we defined 27 equally wide spectroscopic channels between $\lambda=1.12\,\mu$m and $\lambda=1.66\,\mu$m. In the light curve fitting (Stage 30), we discarded the first exposures from each orbit as they are typically affected by strong systematics. We keep the first orbits of all visits, but only analyze the last seven exposures from the first orbits as the earlier exposures show strong systematics ramps that strongly differ from the initial ramps of the remaining orbits of a visit.

The model we used to fit the light curves in Stage 30 includes a linear systematics baseline (with a constant coefficient $c$ and a linear one $v$), an exponential ramp in each orbit (with coefficients $r_1$, $r_2$ and $r_3$, where $r_3$ is only used in the first orbit of each visit), a scaling factor between the forward and backward-scanned exposures and a transit model implemented using the \texttt{batman} package \citep{batman}. The systematics and astrophysical components of the model are identical to the ones in \cite{kreidberg14} and were multiplied to calculate the full model. In the transit model, we fixed HD\,191939\,b's orbital period to 8.8803256 days, its orbital inclination to $88.1^\circ$, the semi-major axis relative to the star's radius to $18.36$ \citep{orellmiquel23} as well as its argument of periastron to $90^\circ$ and the orbital eccentricity to $0$. For the star, we adopt a quadratic limb darkening law for which we fix the coefficients $u_1$ and $u_2$ to model values calculated using \texttt{exotic-LD} \citep{exoticld}, employing the \texttt{stagger} grid \citep{staggergrid}. We fit both the white and the spectroscopic light curves in a two-step process: Firstly, we ran a least squares fit and then sampled the fit parameters' posterior probabilities using the Markov-chain Monte Carlo (MCMC) package \texttt{emcee} \citep{emcee}. We initiated the MCMC walkers in a Gaussian ball centered at the best-fit parameter values from the least squares fit and with a width of a tenth of the parameter uncertainties. In the MCMC, we additionally sample a multiplicative factor $f$ for the data's uncertainties and use 100 walkers which we run for 5000 steps each, discarding the first 2000 steps as burn-in. In all fits, we keep all astrophysical parameters except for the transit midtimes $T_{\text{mid}}$ identical in all visits and the systematics parameters different between visits. We use uninformative priors for all fit parameters except for $r_3$ for which we use a uniform prior with lower and upper edges at -10 and 10, respectively, to prevent it from running off toward unphysical values.

\begin{figure}[htbp!]
	\centering
	\includegraphics[width=\hsize]{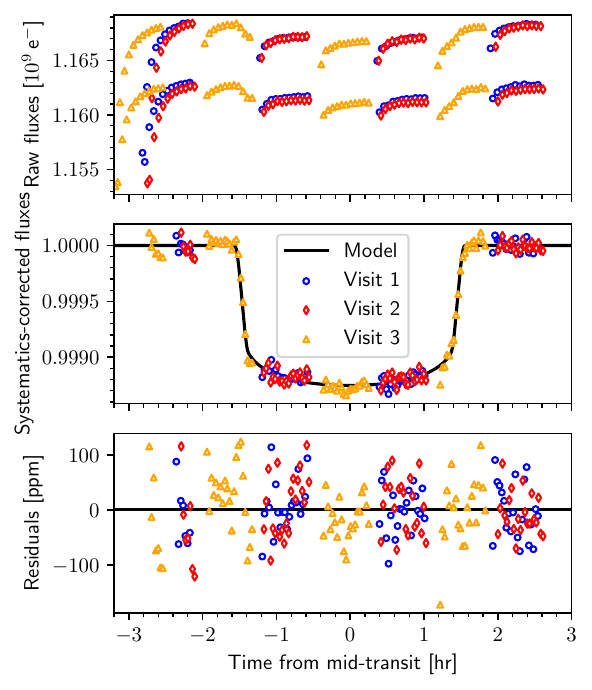}
	\caption{White light curves of HD\,191939\,b's primary transits observed with HST/WFC3 and reduced and modeled with \texttt{PACMAN}. The systematics-corrected light curve only shows the exposures that were analyzed in the light curve fit. A reference line at zero was added in the residuals panel to guide the eye.}
	\label{fig:white-lightcurve}
\end{figure}

In the modeling of the white light curve, we fit for the transit midtimes, the planet's radius relative to the star's radius ($R_p/R_*$), the linear baseline parameters $c$ and $v$, the three exponential ramp parameters $r_1$, $r_2$ and $r_3$, the scaling factor between forward and backward scans and the multiplicative uncertainty factor $f$. The data and maximum-likelihood model are shown in Figure \ref{fig:white-lightcurve} and the root mean square (RMS) of the residuals between the two is 51.6\,ppm. For comparison, the median of the photon noise per exposure of the raw white light curve is 29.3\,ppm.

\begin{figure*}[htbp!]
	\centering
	\includegraphics[width=\hsize]{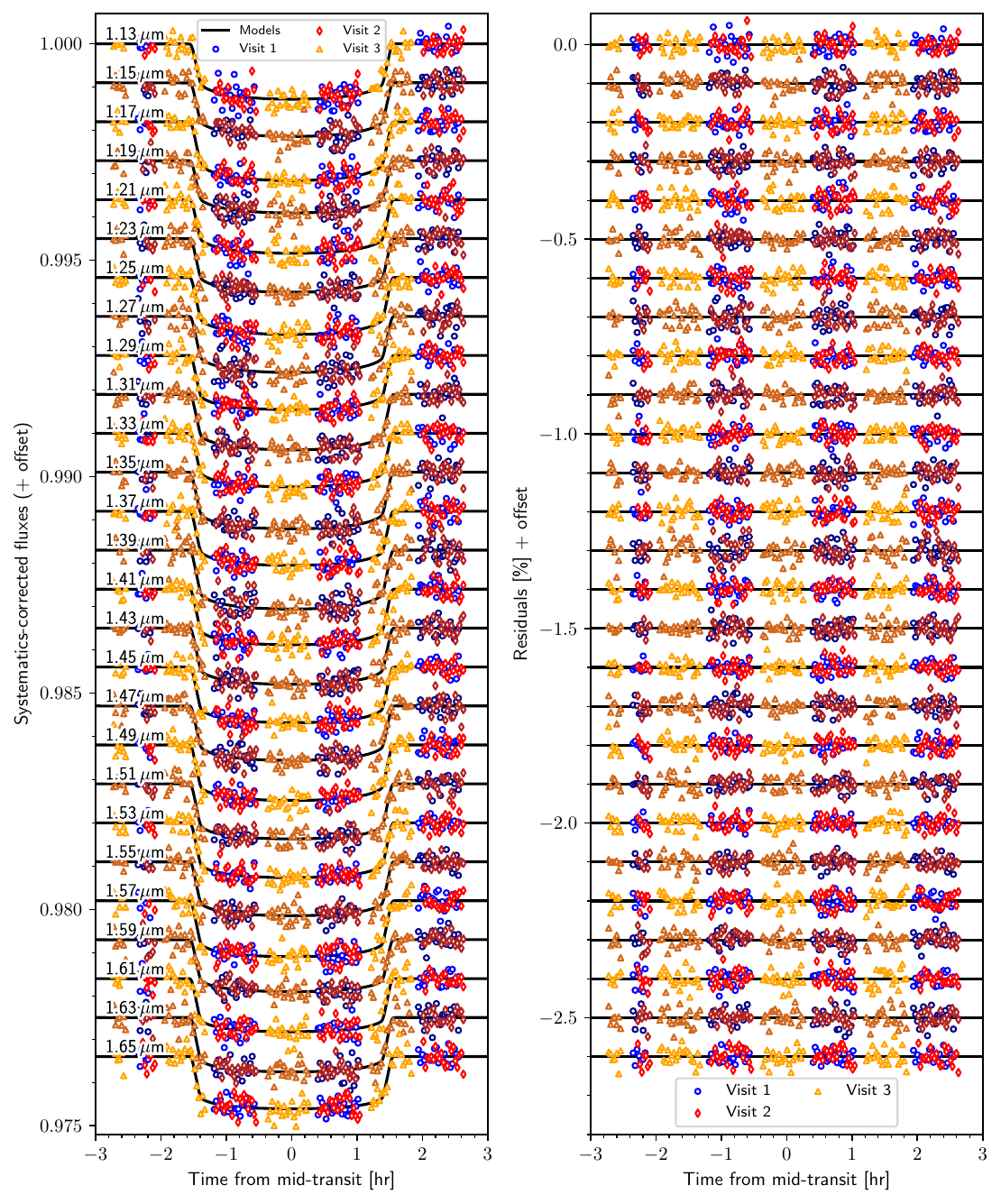}
	\caption{Spectroscopic light curves and fit models with the corresponding residuals from the \texttt{PACMAN} reduction. The data are shown with alternating light and dark colors for visual clarity. Solid black lines at zero were added in the residuals' panel for reference.}
	\label{fig:spectroscopic-lightcurves}
\end{figure*}

\begin{figure}[htbp!]
	\centering
	\includegraphics[width=\hsize]{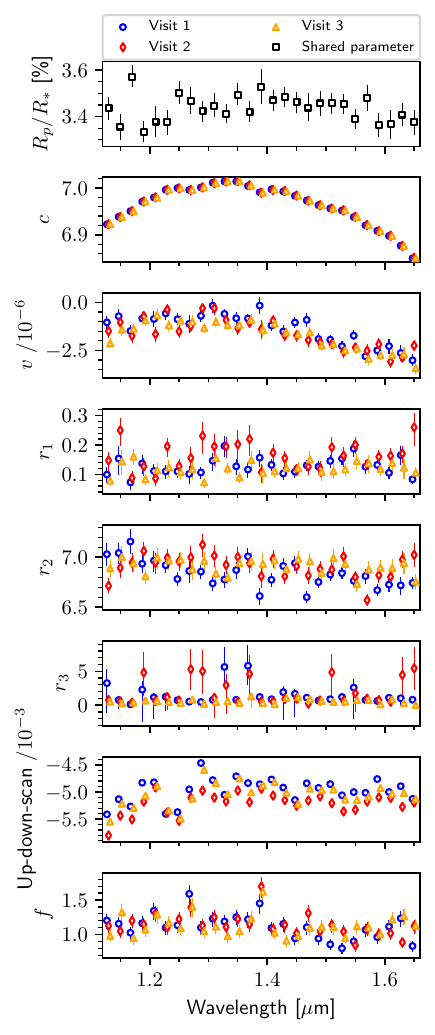}
	\caption{Posteriors of the parameters of the spectroscopic light curve fits from the \texttt{PACMAN} reduction inferred using MCMCs. Black squares show the posteriors of parameters that are employed for all visits' data and colored circles, diamonds and triangles depict posterior medians for parameters that are used for one of the visits only.}
	\label{fig:fit-parameters}
\end{figure}

To fit each spectroscopic light curve, we fixed the transit midtimes to the median values of the white light curve fit and fit for all other model parameters that were allowed to vary in the white light curve fit (i.e., $R_p/R_*$, $c$, $v$, $r_1$, $r_2$, $r_3$, the scale factor between up and down scans and $f$). Figure \ref{fig:spectroscopic-lightcurves} shows the fits to all 27 spectroscopic light curves and Figure \ref{fig:fit-parameters} shows the posteriors of all fit parameters.

\begin{figure}[htbp!]
	\centering
	\includegraphics[width=\hsize]{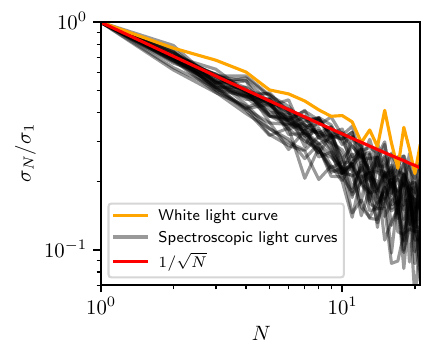}
	\caption{Normalized residual RMS $\sigma_N/\sigma_1$ between best-fit models and data for the white and spectroscopic light curves from the \texttt{PACMAN} reduction as a function of the number of data points $N$ per bin. The expected normalized residual RMS in the absence of red noise \citep{pont06} is shown in red.}
	\label{fig:rms-bins}
\end{figure}

The fits to the spectroscopic light curves are good with the RMS of the residuals between the 27 light curves and models being between $0.85\times$ and $1.44\times$ the photon noise per exposure. In median, the real RMS is $1.06\times$ the photon noise. When binning both the data and models, the residual RMS $\sigma_N$, where $N$ is the number of data points per bin, indicates that there is no time-correlated (`red') noise in the data, since they are generally smaller than $\sigma_1/\sqrt{N}$ (marked with a red solid line in Figure \ref{fig:rms-bins}). In the white light curves, however, there appears to be red noise, since its $\sigma_N$ is generally greater than $\sigma_1/\sqrt{N}$ (see Figure \ref{fig:rms-bins}). Visit 3 stands out with a visible structure in the residuals of the white light curve fit, especially in the third orbit from about 0.5\,hr before until 0.5\,hr after the transit mid-time (see Figure \ref{fig:white-lightcurve}). That data suggest lower fluxes than the best-fit model obtained from fitting all three visits and thus appears to point toward a larger transit depth. The first analyzed data point of the fourth orbit of the same visit at $t=1.2$\,hr after mid-transit shows an anomalously low flux, too (see Figure \ref{fig:white-lightcurve}). These data likely contribute to the hints of red noise in the white light curve visible in Figure \ref{fig:rms-bins}).

\begin{figure*}[htbp!]
	\centering
	\includegraphics[width=\hsize]{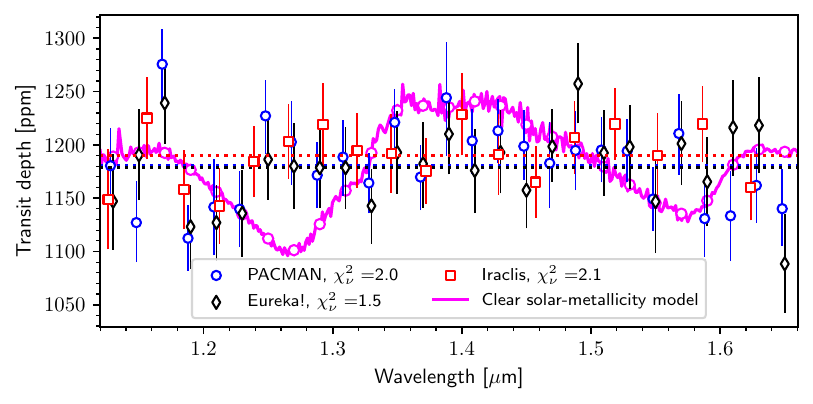}
	\caption{HD\,191939\,b's transmission spectrum observed with HST/WFC3 and reduced with independent reduction pipelines considering all visits and a model spectrum assuming no aerosols in the atmosphere and solar metallicity (see Section \ref{subsec:forward_model}). Colored markers with error bars show the light curve fits' median values and $1\,\sigma$ uncertainties and colored dotted lines depict the data's weighted averages. A small offset in wavelength was subtracted from the data from the \texttt{PACMAN} reduction for visual clarity so that they do not overlap with the \texttt{Eureka!} data points. Magenta circles present the model spectrum binned to the wavelength channels of the \texttt{PACMAN} and \texttt{Eureka!} reductions. The $\chi_\nu^2$ values given in the legend were calculated with each data set and the clear solar-metallicity model.}
	\label{fig:transmission-spectra}
\end{figure*}

\begin{figure*}[htbp!]
	\centering
	\includegraphics[width=\hsize]{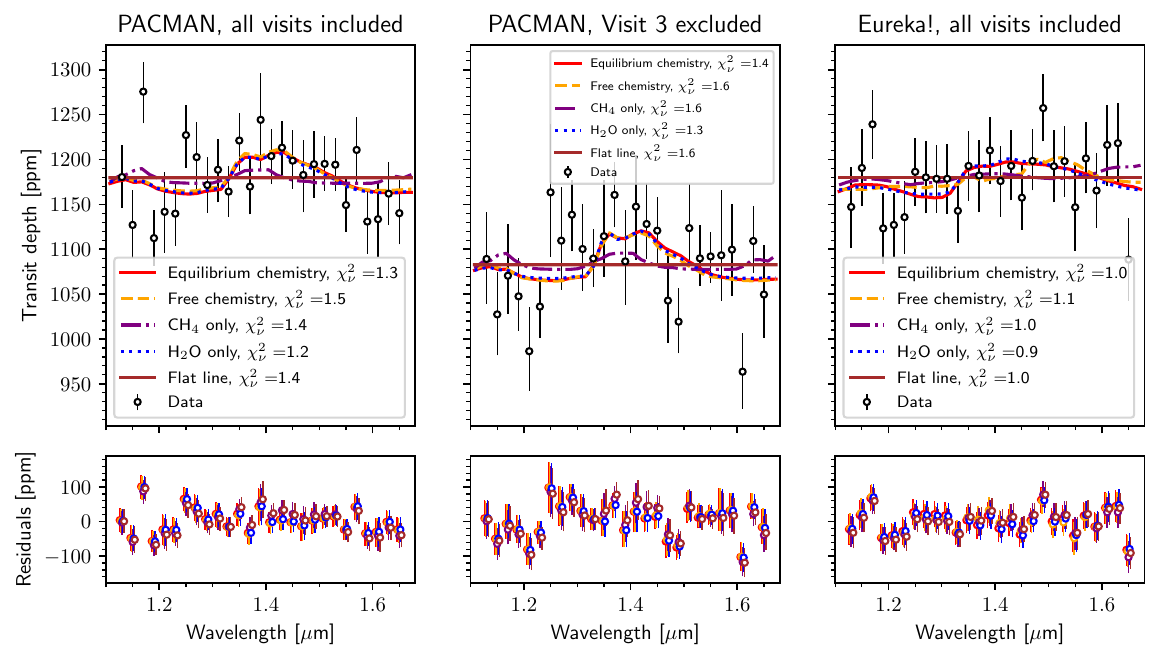}
	\caption{HD\,191939\,b's transmission spectrum obtained using the \texttt{PACMAN} and \texttt{Eureka!} reduction pipelines along with the best-fit model spectra from atmospheric retrievals run with \texttt{petitRADTRANS}. The upper row shows the data and models and the lower row shows the residuals of all models where small wavelength offsets were added to the different models for visual clarity.}
	\label{fig:retrievals}
\end{figure*}

\added{Unlike Visits 1 and 2, Visit 3 samples the in- and egress portions of the transit which are particularly informative about the stellar limb darkening. To investigate whether the limb darkening parameters we fixed to model values in the light curve fits drive the discrepancy between Visit 3 and the other visits, we repeated the light curve fit as before, but additionally fit for the limb darkening. The discrepancy, however, remained unchanged, suggesting that fitting for the limb darkening does not fit the observations more adequately.} To \added{further} examine whether \added{Visit 3} introduces a bias in the final transmission spectrum, we ran the light curve fitting in the same manner as before, but excluded Visit 3 from the analysis. The results from fitting all visits are plotted in Figure \ref{fig:transmission-spectra} and both the results from fitting all visits and when excluding Visit 3 from the light curve fitting are shown in Figure \ref{fig:retrievals}.

\subsubsection{\texttt{Eureka!} reduction}
In parallel to the \texttt{PACMAN} analysis, we reduced and fitted the WFC3 observations of HD\,191939\,b using the \texttt{Eureka!} pipeline \citep{bell22}, employing the \texttt{Eureka!} optimizer \citep{ashtari25} through Stages 1-4 of \texttt{Eureka!}, which include detector-level corrections, background subtraction, spectral extraction, and the generation of light curves. We binned the data to the same 27 spectroscopic light curve channels used in the \texttt{PACMAN} reduction.

For the white-light curve fitting (Stage 5), we performed a joint fit across the light curves from the three visits, with the orbital and transit parameters shared across visits and visit-specific systematics parameters fit independently. Orbital period, inclination, and semi-major axis were fixed \citep{orellmiquel23}; and uniform priors were adopted for the time of transit center, while the planet radius was treated as a shared free parameter across all visits. A quadratic limb-darkening law was used, with both coefficients fitted freely. The systematics model consisted of a linear baseline combined with an exponential ramp, following the default \texttt{Eureka!} implementation for HST/WFC3 observations.

For the spectroscopic light-curve fits, we fixed all shared astrophysical parameters to the posterior medians from the joint white-light curve fit and independently fitted the planet radius in each spectroscopic channel. In contrast to the white light curve analysis, the limb-darkening coefficients were fixed for the spectroscopic fits to wavelength-dependent model values computed from the Stagger 3D stellar grid model \citep{staggergrid}. The three visits were fit separately and we derived HD\,191939\,b's transmission spectrum by computing the weighted mean of the transit depths from all visits for each spectroscopic channel. All other model components and fitting procedures were identical to those used for white light curve fitting. This approach ensures a consistent treatment of the orbital and stellar parameters while allowing wavelength-dependent variations in the transit depth to be captured in the resulting transmission spectrum.

\subsubsection{\texttt{IRACLIS} reduction}
To further validate our dataset analysis, we performed image reduction, light-curve fitting, and 1D spectral extraction using \texttt{IRACLIS} \citep{tsiaras2016b,tsiaras2016a,tsiaras2018}. \texttt{IRACLIS} is an open-source analysis pipeline for reducing and analyzing Hubble Space Telescope (HST) WFC3 IR spectroscopic observations of exoplanet transits and eclipses, specifically single-object spatially scanned datasets taken with the G102 and G141 grisms.

Before extracting the white and spectroscopic light curves, all raw frames were reduced following the steps described in \cite{tsiaras2016a}. We used optimal extraction to obtain the white and spectroscopic light curves. Instrumental systematics (commonly referred to as “ramps”) are known to affect the WFC3 IR detector in both staring and scanning modes. As is common practice in WFC3 data analyses, we discarded the first orbit because it exhibits stronger systematics than subsequent orbits \citep{damiano2017,tsiaras2018}. We fitted the ramps on the white light curve using an approach similar to \cite{kreidberg14}; i.e., we adopted an analytic function with two ramp components, short-term and long-term, to correct the data.

To model the transit light curve, we used \texttt{PyLightcurve} \citep{tsiaras2016a}, which returns the flux as a function of time using the nonlinear limb-darkening law \citep{claret2000}. The limb-darkening coefficients were derived from the profile of a star similar to HD~191939 using the \texttt{PHOENIX} models \citep{husser2013}. The resulting limb-darkening coefficients were fixed and used when fitting both the white and spectroscopic light curves. First, we fitted the white light curve, allowing the mid-transit time, relative planetary radius, and systematics parameters to vary. We then adopted the best-fit mid-transit time when fitting the spectroscopic light curves and computed the 1D spectrum for each visit individually. Finally, we combined the three 1D transmission spectra using a weighted average to obtain the final 1D transmission spectrum shown in Figure~\ref{fig:transmission-spectra}.

\subsection{HST/STIS observations} \label{sec:uv}
The obtained STIS spectra cover the wavelength ranges 1160--1715\,\AA\ (G140L), 1600--3160\,\AA\ (G230L) and the \ion{H}{1}\,1215.67\,\AA\ Lyman\,$\alpha$ line (G140M) with low to medium resolution. The data were retrieved from MAST and automatically reduced with \texttt{calSTIS v3.4.2}. We found that the automated pipeline failed to identify the spectral trace in the G140L and G140M spectra, so we re-extracted the data using \texttt{stistools}\footnote{\url{https://stistools.readthedocs.io/en/latest/}}. Multiple emission lines are visible in the G140L spectrum but they are relatively weak compared with, for example, the Sun, suggesting a low level of chromospheric emission in HD\,191939\,A.
No emission is seen at the \ion{Mg}{2}\,2800\,lines in the G230L spectrum. The weak UV emission is consistent with the long estimated stellar rotation period \citep{lubin24}.

\begin{figure*}[htbp!]
    \centering
    \includegraphics[width=\linewidth]{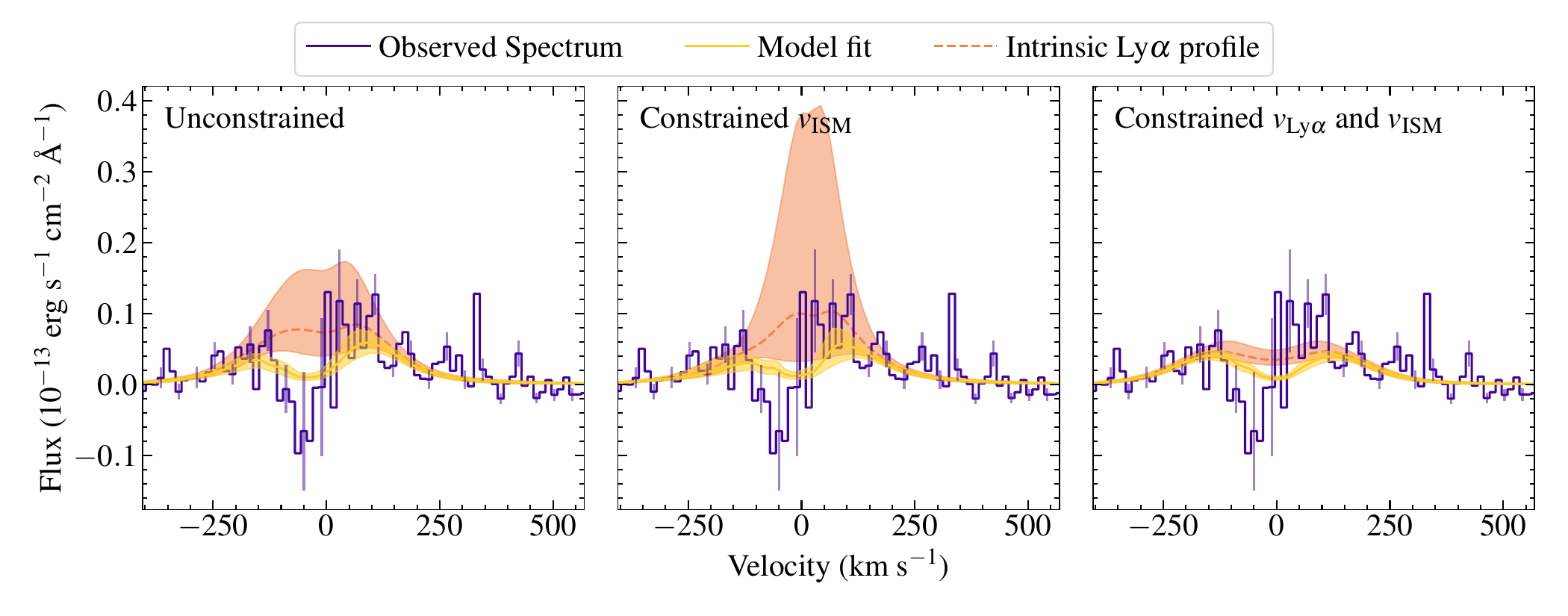}
    \caption{Reconstruction of the Lyman\,$\alpha$ line from the HST/STIS G140M data. All spectra were shifted to the peak of the Lyman\,$\alpha$ line. Purple solid lines with error bars present the observations that show clear absorption of the Lyman\,$\alpha$ peak in line with ISM absorption. The yellow solid lines show the models fit to the observations and the ISM absorption-corrected spectrum of HD\,191939\,A is shown with dashed orange lines. The three versions of the reconstruction using the same data are shown in the different panels.}
    \label{fig:reconstruction}
\end{figure*}

In line with the weakness of HD\,191939\,A's UV emission lines in general, the wings of the Lyman\,$\alpha$ line around the interstellar medium (ISM) absorption are only measured with a low signal-to-noise ratio in the G140M spectrum. No other features are detected in the G140M spectrum. Following \cite{lyapy}, we reconstructed the full Lyman\,$\alpha$ line emitted by HD\,191939\,A, fitting for the integrated Lyman\,$\alpha$ flux $F_{\text{Ly}\alpha}$, the \ion{H}{1} column density $N($\ion{H}{1}$)$, the Doppler velocity of the Lyman\,$\alpha$ line $v_{\text{Ly}\alpha}$ and the Doppler velocity of the absorbing local ISM (LISM) $v_{\text{LISM}}$. The results for each parameter's posterior are listed in Table \ref{tab:lya} and the data together with the models and reconstructed LISM absorption-corrected spectra are shown in Figure \ref{fig:reconstruction}. When sampling these reconstruction parameters with uninformative priors (`Unconstrained fit' column in Table \ref{tab:lya}), we find an $N($\ion{H}{1}$)$ that is consistent with the LISM at HD\,191939\,A's distance of 53.91\,pc (\citealp{youngblood25} list $\log_{10}(N($\ion{H}{1}$)/\text{cm}^{-2})=18.4$ as a typical value for distances between 20 and 70\,pc) and a $v_{\text{Ly}\alpha}$ that is in line with HD\,191939\,A's radial velocity of $-(9.52\pm 0.19)$\,km\,s$^{-1}$ \citep{gaia_dr2}. $v_{\text{LISM}}$, however, is discrepant with the value of zero adequate for the local interstellar cloud (LIC, \citealp{redfield08}) by $2.7\,\sigma$. To validate the reconstructed value for $F_{\text{Ly}\alpha}$, we ran the reconstruction again, constraining $v_{\text{LISM}}$ with a Gaussian prior of $(-0.48\pm1.38$)\,km\,s$^{-1}$ based on \citet{redfield08} (`Constrained $v_{\text{LISM}}$' column in Table \ref{tab:lya}). In this approach, we found a value of $F_{\text{Ly}\alpha}$ that is consistent with the unconstrained fit, but with a larger uncertainty. \added{The median value of $v_{\text{Ly}\alpha}$, though, moves away from the star's systemic velocity measured with Gaia.} Thus, finally, we ran a third reconstruction, with the same constraint on $v_{\text{LISM}}$ and using a Gaussian prior for $F_{\text{Ly}\alpha}$ with the Gaia values, which returned a $F_{\text{Ly}\alpha}$ consistent with the two previous reconstruction runs (see `Constrained $v_{\text{LISM}}$ and $v_{\text{Ly}\alpha}$' column in Table \ref{tab:lya}). We thus conclude that the reconstructed flux is robust against choices of freely sampling or constraining $v_{\text{LISM}}$ and $v_{\text{Ly}\alpha}$ using literature values.
\begin{table*}[htbp!]
    \centering
    \caption{Key parameters of the reconstructed Lyman\,$\alpha$ line.}
    \begin{tabular}{lrrr}\hline \hline
         Parameter &  Unconstrained fit & Constrained $v_{\text{LISM}}$ & Constrained $v_{\text{LISM}}$ and $v_{\text{Ly}\alpha}$ \\ \hline
         $F_{\text{Ly}\alpha}$ [$\text{erg}\,\text{s}^{-1}\,\text{cm}^{-2}$] & $1.3^{+1.0}_{-0.5} \times10^{-14}$ & $1.3^{+2.0}_{-0.6} \times10^{-14}$ & $8.7^{+2.6}_{-1.8} \times10^{-15}$\\
         $\log_{10}$(N(\ion{H}{1}[cm$^{-2}$]) & $18.0^{+0.4}_{-0.6}$ & $17.9^{+0.4}_{-0.6}$ & $17.6^{+0.5}_{-0.4}$\\
         $v_{\text{Ly}\alpha}$ [km\,s$^{-1}$] & $-11^{+17}_{-13}$ & $12^{+16}_{-24}$ & $-9.2\pm0.2$ \\
         $v_{\text{LISM}}$ [km\,s$^{-1}$] & $-38^{+14}_{-9}$ & $-0.7\pm1.4$ & $-0.7\pm1.4$\\ \hline
    \end{tabular}
    \label{tab:lya}
\end{table*}

\begin{figure*}[htbp!]
    \centering
    \includegraphics[width=\linewidth]{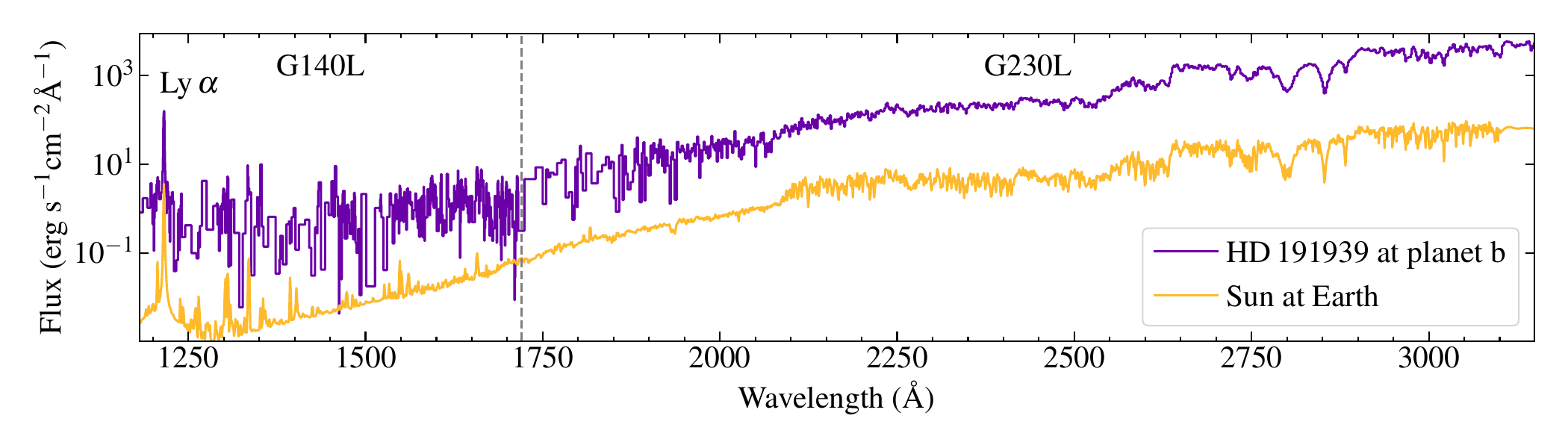}
    \caption{STIS spectra of HD\,191939\,A combined with the Lyman\,$\alpha$ reconstruction. The spectrum is scaled to the distance of HD\,191939\,b and compared with the quiet Sun from \cite{woods09}. The STIS spectrum has been downsampled to remove negative flux points.}
    \label{fig:stis}
\end{figure*}

With the three performed reconstructions agreeing in the reconstructed Lyman\,$\alpha$ fluxes, we derive HD\,191939\,A's spectrum around the Lyman\,$\alpha$ line from the unconstrained fit to propagate possible measurement uncertainties to the final spectrum. The spectrum combined from the HST/STIS observations, including the reconstructed Lyman\,$\alpha$ line is shown in Figure \ref{fig:stis}.

\subsection{Host star photometric monitoring} \label{sec:monitoring}

We conducted photometric monitoring of the host star, HD\,191939\,A, with the automated 24-inch telescope at Van Vleck Observatory, Wesleyan University in Connecticut, USA, as part of the same host star photometric monitoring campaign presented in \cite{kahle25}. We observed HD\,191939\,A from March 2023 to April 2024, which covers Visit 3 of the three HST transit observations. To construct the dataset, we require at least four valid exposures on a given observing night in each of the BVRI bands separately. We also visually identify cloud passing conditions and exclude the corresponding nights. Consequently, we utilize 1294, 1359, 1282, 1212 individual exposures from 80, 84, 80, 81 observing nights in the BVRI bands, respectively. The median observation cadence is between once every two days and once every three days, while the mean value is almost five days.

We follow the same data reduction procedure as in \cite{kahle25} for plate solving, image calibration, and aperture photometry, but implement a different relative photometry routine. We select field stars that 1) are detected in at least 89\,\% of the total exposures, and 2) exhibit no obvious brightness fluctuation, as comparison stars. This results in two selected stars. We note the presence of fluctuations in field stars not selected for photometric comparison, but have ensured that the two comparison stars are free from fluctuations. We verified the robustness of the selection by testing with other comparison star combinations, and the resulting light curves did not change significantly.

\begin{figure*}[htbp!]
	\centering
	\includegraphics[width=\hsize]{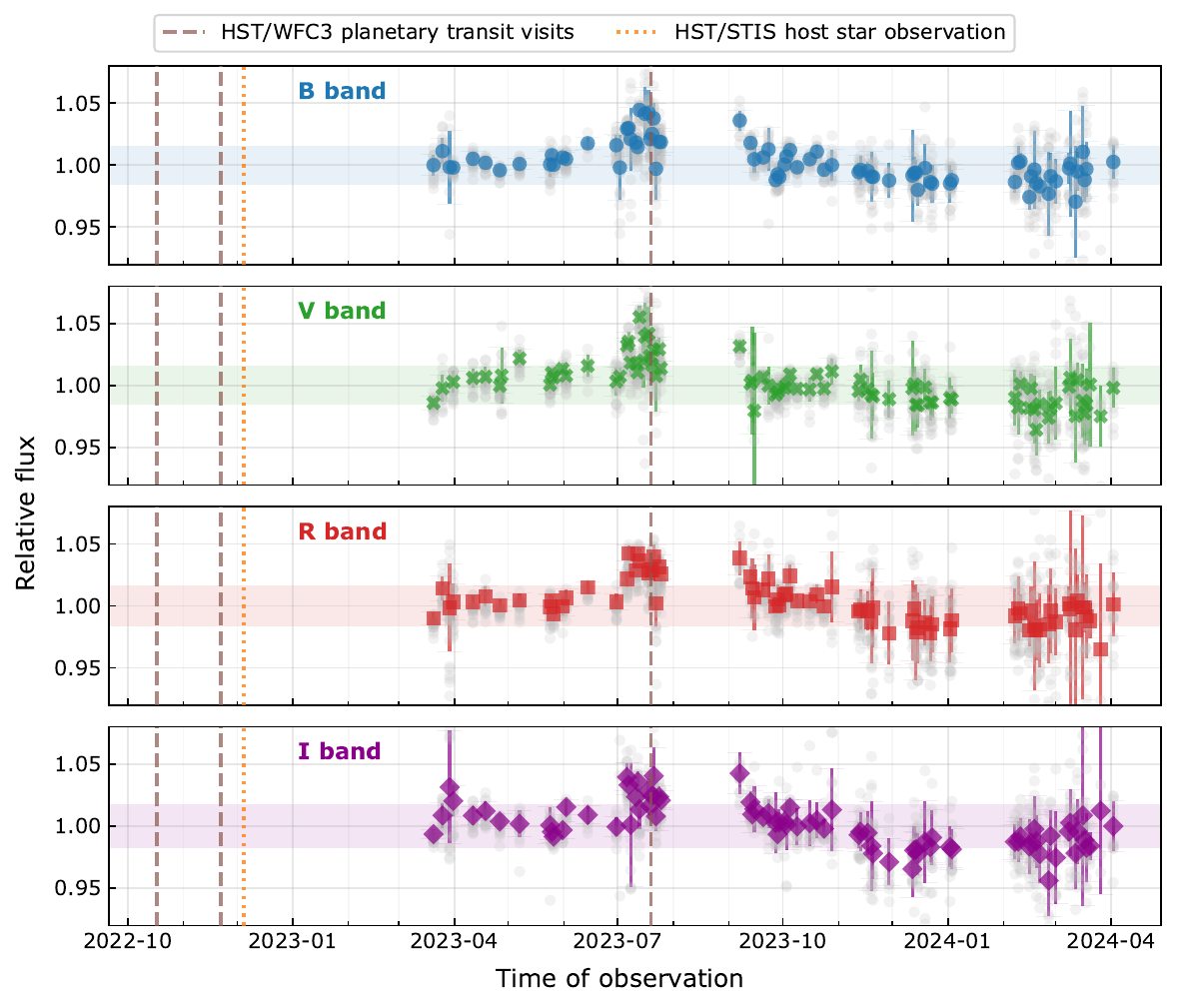}
	\caption{Light curves of HD\,191939\,A in the BVRI bands observed with the automated 24-inch telescope at Van Vleck Observatory, Wesleyan University. Binned daily photometry is shown with large, colored dots. Each error bar represents the standard deviation of the underlying individual exposures (small, gray dots) within a night. The shaded areas show the $1\sigma$ ranges of the binned photometry.}
	\label{fig:monitoring}
\end{figure*}

\begin{figure}[htbp!]
	\centering
	\includegraphics[width=\hsize]{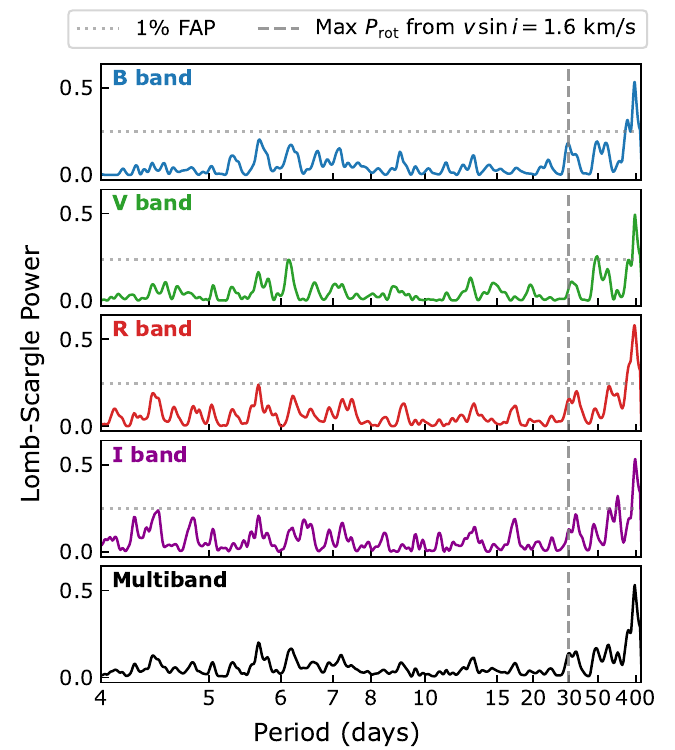}
	\caption{Lomb-Scargle periodograms of the binned daily photometry. We also display the corresponding 1\% false-alarm probability (FAP) levels in the BVRI bands and an estimated upper limit of HD\,191939\,A's rotational period from $v \sin{i} = (1.6 \pm 0.3)\,\text{km/s}$ \citep{lubin24}.}
	\label{fig:periodogram}
\end{figure}

In our final light curve, presented in Figure \ref{fig:monitoring}, HD\,191939\,A seems to exhibit brightness fluctuations on the time scale of months with a pronounced brightness maximum in July 2023 when Visit 3 of the HST/WFC3 observations was taken. In Figure \ref{fig:periodogram}, we attempt to search for possible periodicity with Lomb-Scargle periodograms in BVRI bands independently and through a multiband implementation in Astropy \citep{vdp15, astropy22}. We scrutinize peaks that yield a low false-alarm probability and find that some of them are effects of the window function and potential aliasing, while the peak between 300 days and 400 days is likely an artifact of annual or seasonal cycles, which is common in ground-based datasets \citep{vdp18}. We also note that the long period peaks would not be consistent with the $v \sin{i}$ measurement \citep{lubin24}. Although we cannot determine if the variability is periodic, we deem the variability of HD\,191939\,A to be real, possibly related to its long, undetermined rotation. We find no simultaneous flux increase in the four bands and therefore determine that there is no evidence of stellar flares. 

\section{Atmospheric analysis}
To quantify possible constraints on HD\,191939\,b's atmospheric composition, we analyzed its transmission spectrum measured with HST/WFC3 using the atmospheric modeling framework \texttt{petitRADTRANS} \citep{prt1,prt2}. In all models, we use Rayleigh opacities for molecular hydrogen (H$_{2}$) and helium (He), and continuum opacities from H$_{2}$-H$_{2}$ and H$_{2}$-He as well as line opacities for H$_{2}$O \citep{h2o_opacities}, CO$_{2}$ \citep{co2_opacities}, CO \citep{co_opacities}, CH$_{4}$ \citep{ch4_opacities} and NH$_{3}$ \citep{nh3_opacities}. All line opacities and all computed models have a spectral resolution of $R=600$.

\subsection{Null hypothesis tests} \label{subsec:forward_model}
To quantify the information content of the data, we generated a model transmission spectrum of HD\,191939\,b assuming an aerosol-free atmosphere using an isothermal temperature profile of 880\,K, a stellar radius of $0.94\,R_\odot$ \citep{lubin22}, a planetary radius of $3.41\,R_\oplus$, a planetary mass of $10.0\,$M$_\oplus$ \citep{orellmiquel23} and a reference pressure of 10\,mbar. We assume that the abundances of chemical species are set according to a network of chemical reactions (referred to as `equilibrium chemistry') and adopt solar metallicity and a carbon-to-oxygen ratio (C/O) of 0.55 for elemental abundances. While it is unlikely that these assumptions are true for HD\,191939\,b, the resulting model primarily serves for estimating the amplitude of absorption features of some planetary atmosphere to estimate how constraining the acquired observations actually are.

For each data reduction, we add a constant offset to the model so that the model mean and the weighted means of the data are equal and calculate the $\chi^2$ and number of degrees of freedom $\nu=n-1$, where $n$ is the number of data points. From these, we calculate the p-value of each data set and convert it into a frequentist significance $N\sigma$ (see \cite{Gregory05}, p. 162ff). We find that the \texttt{PACMAN} reduction is inconsistent with the clear solar-metallicity atmosphere in chemical equilibrium at $3.2\,\sigma$ when fitting all visits and at $2.4\,\sigma$ when excluding Visit 3. The \texttt{Eureka!} reduction is inconsistent at $2.0\,\sigma$ and the \texttt{Iraclis} reduction at $2.7\,\sigma$.

We then performed another set of hypothesis tests with the transit depth as a constant in wavelength and thus a transmission spectrum without absorption features. In the same procedure as before, the \texttt{PACMAN}, \texttt{Eureka!} and \texttt{Iraclis} reductions are inconsistent with the null hypothesis at $1.3\,\sigma$, $0.4\,\sigma$ and $0.08\,\sigma$, respectively, when fitting all visits. When Visit 3 is excluded from the \texttt{PACMAN} reduction, the transmission spectrum is inconsistent with a flat line at $1.5\,\sigma$.

\subsection{Atmospheric retrievals} \label{sec:retrievals}
We ran retrievals on the \texttt{PACMAN} reduction including and excluding Visit 3 and the \texttt{Eureka!} reduction to explore the impact of data-reduction assumptions and possible systematic differences introduced by Visit 3. We applied five different model setups: Firstly, an equilibrium chemistry setup with a metallicity parameter and the C/O as free parameters. Secondly, a `free chemistry' setup with vertically-constant abundances of H$_2$O, CO, CO$_2$ and NH$_3$ that are fit for individually. Thirdly, an H$_2$O-only setup which is identical to the free chemistry setup but excludes all molecules apart from H$_2$O. Fourthly, a CH$_4$ setup which is the same as the H$_2$O scenario but with atmospheric CH$_4$ rather than H$_2$O. And finally, a flat line which includes no molecular opacities. We adopt an isothermal pressure-temperature profile with temperature $T_{\rm iso}$. For atmospheric aerosols, we include a grey cloud deck with the cloud top pressure as a free parameter. Below that cloud top pressure, the atmosphere is opaque in all wavelengths and thus, this model corresponds to either cloud or haze particles abundant enough in appropriate particle sizes to make the atmosphere below the cloud top pressure completely grey in the observed wavelengths. Table \ref{tab:atmospheric_retrieval_priors} summarizes the priors used in the retrievals.
\begin{table*}[htbp!]
    \centering
    \caption{Parameter priors in the atmospheric retrievals.}
    \label{tab:atmospheric_retrieval_priors}
    \begin{tabular}{lccc}
    \hline\hline
    Parameter & Description & Prior & Retrieval setup  \\ \hline
    $T_{\rm iso}$ [K] & Isothermal temperature & $\mathcal{U}(0,2000)$ & All \\
    $R_{\rm p}$ [R$_\oplus$] & Planet radius & $\mathcal{N}(3.39,0.07)$ & All \\
    $M_{\rm p}$ [M$_\oplus$] & Planet mass & $\mathcal{N}(10.4,0.9)$ & All \\
    log($P_{\rm ref}$) [bar] & Reference pressure & $\mathcal{U}(-6,3)$ & All \\
    log($P_{\rm cloud}$) [bar] & Cloud top pressure & $\mathcal{U}(-6,3)$ & All \\
    log(M/H) [$\times$ solar] & Atmospheric metallicity & $\mathcal{U}(-1,3)$ & Equilibrium only \\
    C/O & Carbon-to-Oxygen ratio & $\mathcal{U}(0.1,1.6)$ & Equilibrium only \\
    log($X_{species}$) & Molecular mass fractions & $\mathcal{U}(-10,0)$ & Free chemistry only \\
    \hline
    \end{tabular}
    \tablecomments{In the prior column, $\mathcal{U}$ represents a uniform prior with the lower and upper edges given in parentheses. $\mathcal{N}$ refers to a Gaussian prior with the mean and standard deviation listed in parentheses. The priors for $R_{\rm p}$ and $M_{\rm p}$ were adopted from the results of \cite{lubin22}}
\end{table*}

None of the atmospheric retrieval setups delivers a clear Gaussian peak in the posterior probabilities of any parameter governing the chemical composition of HD\,191939\,b's atmosphere in any reduction of the transmission spectrum. The best-fit spectra from all atmospheric retrievals are shown in Figure \ref{fig:retrievals} along with the data and show the reason for the lack of constraints on the planet's atmospheric composition: There is no clear molecular absorption feature in the data, thus giving a flat spectrum. For a quantitative comparison between the investigated atmospheric scenarios, we show the natural logarithm of their Bayes factors ($\ln(\mathcal{B})$) with respect to the flat line model in Table \ref{tab:bayes_atm_retrievals}. In all reductions of the transmission spectrum, none of the tested atmospheric models is statistically significantly preferred over the flat line model, since all $\ln\mathcal{B} < 1.15$ which corresponds to evidence `barely worth mentioning' \citep{thorngren26}.
\begin{table}[htbp!]
    \centering
    \caption{Bayesian evidences $Z$ for our suite of atmospheric retrievals and derived Bayes factors $\mathcal{B}$ with respect to the flat line model.}
    \label{tab:bayes_atm_retrievals}
    \begin{tabular}{llcc}
    \hline\hline
    Reduction & Retrieval                      &  $\ln(Z)$   &   $\ln(\mathcal{B})$   \\ \hline
    \texttt{PACMAN} & flat line                         &  232.5   &   -   \\
    \texttt{PACMAN} & H$_{2}$O only                          &  233.4   &   0.90   \\
    \texttt{PACMAN} & CH$_{4}$ only                          &  232.1   &  -0.40    \\
    \texttt{PACMAN} & equilibrium                       &  233.1   &  0.60    \\
    \texttt{PACMAN} & free                              &  231.5   &  -1.0    \\
    \texttt{PACMAN} (Visit 3 excluded) & flat line              &  224.4   &  -    \\
    \texttt{PACMAN} (Visit 3 excluded) & H$_{2}$O only               &  225.5    &  1.10      \\
    \texttt{PACMAN} (Visit 3 excluded) & CH$_{4}$ only               &  224.3    &  -0.10     \\
    \texttt{PACMAN} (Visit 3 excluded) & equilibrium            &  224.9    &  0.50      \\
    \texttt{PACMAN} (Visit 3 excluded) & free                   &  223.7    &  -0.70     \\
    \texttt{Eureka!} & flat line                         &  233.5   &      -   \\
    \texttt{Eureka!} & H$_{2}$O only                          &  233.6   &  0.10    \\
    \texttt{Eureka!} & CH$_{4}$ only                          &  233.1   &  -0.40    \\
    \texttt{Eureka!} & equilibrium                       &  233.4   &  -0.10    \\
    \texttt{Eureka!} & free                              &  232.6   &  -0.90  \\
    \hline
    \end{tabular}%
\end{table}

\begin{figure*}[htbp!]
    \centering
    \includegraphics[width=\linewidth]{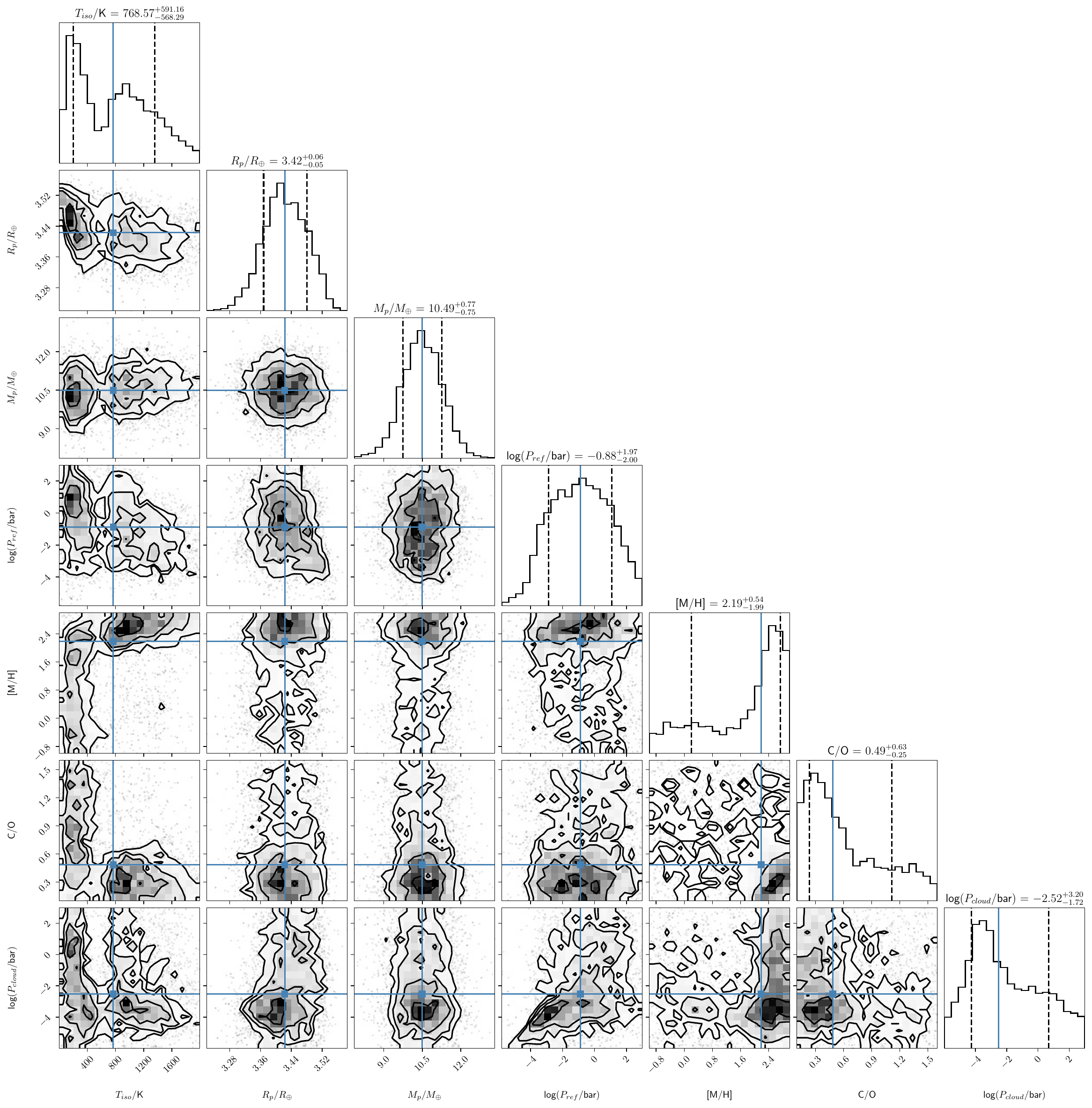}
    \caption{Posterior probabilities of the equilibrium chemistry retrieval on the \texttt{PACMAN} reduction considering all visits. Solid blue lines indicate the samples' medians and dashed black lines show the 16th and 84th percentiles.}
    \label{fig:corner}
\end{figure*}

The posterior probabilities of the retrievals reveal a cloud-metallicity degeneracy (see, e.g., \citealp{benneke13}): High-altitude aerosols, an atmosphere with a high mean-molecular weight or a combination of both of these scenarios can fit the data (see Figure \ref{fig:corner}). Metallicities ranging from sub-Solar to several hundred times Solar and cloud top pressures lower than $\sim 1\,$bar are consistent with the observations within $1\sigma$.

\section{Discussion}
\subsection{Comparison of HST/WFC3 data reductions}
We reduced the HST/WFC3 data using three independent approaches that employed the open-source pipelines \texttt{PACMAN}, \texttt{Eureka!} and \texttt{Iraclis} to examine if the different analysis choices impact the derived transmission spectrum of HD\,191939\,b. One key difference between the different approaches is that \texttt{PACMAN} starts the reduction at images calibrated by MAST, whereas \texttt{Eureka!} and \texttt{Iraclis} perform calibration steps themselves on the raw images. In \texttt{Eureka!}, we fit the white light curves of all three visits jointly and the spectroscopic light curves visit-per-visit before calculating a weighted average. In \texttt{Iraclis}, we fit all data visit-by-visit. Despite these differences in reduction approaches, the analyses overall give compatible results with all transmission spectra being consistent with a flat line and delivering moderate evidence between $2.0$ and $3.2\,\sigma$ against a cloud-free atmosphere with solar metallicity. The agreement between the \texttt{PACMAN} and \texttt{Eureka!} reductions that share the same wavelength bins, is especially good since all data points have overlapping error bars (see Figure \ref{fig:transmission-spectra}). The difference in rejection significance of the cloud-free solar metallicity atmosphere despite the close agreement of the \texttt{PACMAN} and the \texttt{Eureka!} spectra is possibly caused by the latter having larger uncertainties than the former: \texttt{PACMAN}'s median uncertainty on the transit depths is 34.7\,ppm per channel, while \texttt{Eureka!}'s median uncertainty is 39.3\,ppm.

\subsection{Stellar activity}
As the inspection of the white light curves analyzed with \texttt{PACMAN} suggested (see Section \ref{subsec:pacman}), the overall transit depth of the \texttt{PACMAN} spectrum decreases when Visit 3 is discarded (see Figure \ref{fig:retrievals}). This \added{transit depth offset} could be driven by unocculted star spots during that observation that make the surface of the star occulted by the planet anomalously bright with respect to the whole disc, giving a planet that appears larger. \added{This brightness difference between HD\,191939\,b's transit chord and the whole disk could additionally be compounded by bright faculae in the transit chord.} The possible presence of star spots \added{and faculae} is supported by HD\,191939\,A's UV spectrum (see Section \ref{sec:uv}) that includes chromospheric emission lines and the photometric monitoring that reveals brightness variations (see Section \ref{sec:monitoring}), both lines of evidence pointing toward stellar activity. During Visit 3 that is possibly affected by star spots \added{and faculae}, HD\,191939\,A is close to a brightness maximum in the monitored epoch (see Figure \ref{fig:monitoring}). In the Sun, high levels of activity that drive larger numbers of Sun spots correlate with an increased total brightness (see, e.g., \citealp{frohlich13}). Thus, since HD\,191939\,A, like the Sun, is a G-type star, the higher brightness during Visit 3 might indicate \added{increased levels of stellar activity} during that HST/WFC3 observation. For HD\,191939\,A, \added{a} correlation between the total brightness and \added{activity levels} is only hypothetical, though, and we cannot verify this idea further due to the lack of photometric monitoring coverage during Visits 1 and 2.

\subsection{Possible atmospheres in HD\,191939\,b}
Independent of the applied data reduction pipeline and of including or excluding Visit 3 from the data reduction, there is no statistically significant evidence for absorption features in the data (see Section \ref{sec:retrievals}). Thus, the interpretation of HD\,191939\,b’s atmospheric composition is limited by its transmission spectrum lacking discernible features. Comparing the data with a forward model delivers rejection significances between $2.0\,\sigma$ and $3.2\,\sigma$ against the cloud-free atmosphere with solar metallicity depending on the data reduction. Thus, the observations are not constraining enough to conclusively rule out that scenario. However, some aerosol cover or an elevated metallicity muting spectral features through a high mean molecular weight and thus a small atmospheric scale height is likely. Aerosols are indeed probably present in HD\,191939\,b, since both hydrocarbon hazes formed through photochemistry \citep{zahnle09,morley15} and clouds \citep{morley15,helling23,carone25} can accumulate at its equilibrium temperature of $(880\pm20)$\,K. A high mean molecular weight atmosphere might be driven by metal enrichment of the planet compared to solar abundances. This is plausible in HD\,191939\,b due to its mass of $0.032\pm0.002$ Jupiter masses \citep{orellmiquel23}, since both the Solar System planets and exoplanets indicate increasing metal enrichment with decreasing planet mass down to masses of about 0.07 Jupiter masses (see, e.g., \citealp{thorngren16,chachan25}). Indeed, observations with the James Webb Space Telescope (JWST) have confirmed metal-enriched atmospheres in multiple sub-Neptunes \citep{kempton23,roy23,benneke24,piauletghorayeb24,ahrer25,hu25}. A combination of a high mean molecular weight, hazes and clouds in HD\,191939\,b's atmosphere is also possible: The high metal content of a heavy atmosphere can deliver the heavy ingredients necessary for cloud and haze formation, additionally muting atmospheric absorption features.

\subsection{HD\,191939\,b's featureless transmission spectrum in context}
Following the approach of \cite{brande24}, we estimate the size of a hidden H$_2$O absorption feature $A_H$ in HD\,191939\,b by calculating the difference in our model spectra assuming a pure H$_2$O atmosphere between $\lambda=1.4\,\mu$m and $\lambda=1.25\,\mu$m in units of atmospheric scale heights. Also in line with \cite{brande24}, we calculate HD\,191939\,b's equilibrium temperature assuming an albedo of 0.3 which results in $T_{\text{eq}} = 818\,$K when adopting stellar and planetary parameters from \cite{lubin22} and \cite{orellmiquel23}. We note that this value deviates from the zero-albedo $T_{\text{eq}} = (880\pm 20)\,$K quoted in the other parts of this work, however, it is necessary to use $T_{\text{eq}}= 818\,$K here for a direct comparison to the results of \cite{brande24}. Assuming a mean molecular weight $\mu=3.05$ which corresponds to an atmosphere with a metallicity of $100\times$ solar, we find $A_H = 1.19^{+0.62}_{-0.67}$ and $A_H = 1.28^{+0.83}_{-0.82}$ in the retrievals on the \texttt{PACMAN} reduction with and without Visit 3, respectively, and $A_H = 0.72^{+0.63}_{-0.65}$ with the \texttt{Eureka!} reduction. For $T_{\text{eq}}=818$\,K, the second-order polynomial \cite{brande24} fit to previous HST observations returns $A_H=1.99^{+0.27}_{-0.27}$. Thus, HD\,191939\,b’s transmission spectrum delivers an H$_2$O feature that is smaller than what the population analysis suggests, however, it is still consistent within $1.85\,\sigma$.

HAT-P-11\,b, HD\,106315\,c and HAT-P-26\,b that \cite{brande24} used to probe the temperature range between 796\,K and 909\,K have radii of $4.90\,R_\oplus$, $4.38R_\oplus$ and $7.31\,R_\oplus$, respectively \citep{hatp11,hd106315,hatp26}, and are thus not sub-Neptunes that have radii no larger than $4\,R_\oplus$. HIP\,41378\,b, the only sub-Neptune in the temperature range in question considered by \cite{brande24}, only delivered a very poorly constrained measurement of $A_H=2.64^{+2.62}_{-2.50}$ and thus presumably did not contribute significantly to the quadratic fit. Therefore, the discrepancy between HD\,191939\,b's transmission spectrum and the population study of \cite{brande24} might indicate a systematically more efficient muting of absorption features in sub-Neptunes than in giant planets in the temperature range between 796\,K and 909\,K, perhaps driven by an elevated aerosol coverage or atmospheric metallicity. If there is such a hypothetical systematic difference, it appears that it might cease at equilibrium temperatures higher than some value between $818$\,K (HD\,191939\,b's equilibrium temperature assuming an albedo of 0.3) and TOI-421\,b's equilibrium temperature of $(922\pm 14)$\,K \citep{krenn24}, since the transmission spectrum of the sub-Neptune TOI-421\,b closely matches the fit of \cite{brande24} to previous HST observations \citep{davenport25}.

Alternatively, the growing sample of sub-Neptunes that are discrepant with the analysis of \cite{brande24}, GJ\,1214\,b, GJ\,3090\,b, LP791-18\,c \citep{roy25}, and now HD\,191939\,b, likely indicates that the population of sub-Neptunes is diverse beyond a simple relation with equilibrium temperatures. Differences in bulk composition driven by, e.g., formation in the inner part of the protoplanetary disk and formation in the outer part with subsequent migration inwards might drive profoundly different atmospheric chemical compositions and mean molecular weights.

\section{Conclusions}
The transmission spectrum of the sub-Neptune HD\,191939\,b was observed using HST/WFC3 and the host star was characterized both by measuring its UV spectrum with HST/STIS and by monitoring its photometric magnitude with the Van Vleck Observatory. Chromospheric emission is detected in HD\,191939\,A's UV spectrum (see Figure \ref{fig:stis}), though the amplitudes of the corresponding emission lines are consistent with moderate activity levels. In line with stellar activity, the photometric monitoring suggests anomalously high apparent magnitudes and thus possibly the presence of activity during Visit 3 of the HST/WFC3 observations (see Figure \ref{fig:monitoring}) which might affect the data acquired during that visit (see Figures \ref{fig:white-lightcurve} and \ref{fig:retrievals}). Therefore, stellar contamination of the transmission spectra of HD\,191939\,b, c and d observed in the future is possible.

HD\,191939\,b's transmission spectrum observed with HST/WFC3 is featureless, though the data are not constraining enough to exclude the possibility of an aerosol-free atmosphere with solar metallicity at sufficient statistical significance. However, aerosols or a high mean molecular weight of the atmosphere muting spectral features or both of these effects combined, are likely. Comparing HD\,191939\,b's transmission spectrum to the spectra of colder sub-Neptunes and giant planets with similar temperatures as HD\,191939\,b suggests two possible scenarios: Firstly, gas giants between equilibrium temperatures of 796\,K and 909\,K, assuming albedos of 0.3, produce a larger scale-height normalized H$_2$O feature in HST/WFC3's wavelength range than sub-Neptunes. This might be driven by higher metallicities and more aerosol coverage in the latter. And secondly, sub-Neptunes are a diverse planet population whose atmospheres are more complex than a direct consequence of the planets' equilibrium temperatures.

\begin{acknowledgements}
This research is based on observations made with the NASA/ESA Hubble Space Telescope obtained from the Space Telescope Science Institute, which is operated by the Association of Universities for Research in Astronomy, Inc., under NASA contract NAS 5–26555. Q.C.T.~and S.R.~wish to thank Roy Kilgard for his assistance in the operation of the Wesleyan 24-inch telescope and the data storage system. Q.C.T.~offers his heartfelt thanks to Szymon W. Petyniak for the helpful discussions on stellar activity.  K.A.K.~gratefully acknowledges support from the DLR via project P.S.ASTR1508. T.D. acknowledges support from the McDonnell Center for the Space Sciences at Washington University in St. Louis. Part of the research was carried out at the Jet Propulsion Laboratory, California Institute of Technology, under a contract with the National Aeronautics and Space Administration (80NM0018D0004).
\end{acknowledgements}

\section*{Data availability}
\added{The HST data presented in this article are associated with HST program GO 17192 and were obtained from the Mikulski Archive for Space Telescopes (MAST) at the Space Telescope Science Institute. The specific observations analyzed can be accessed via \dataset[DOI: 10.17909/gxjc-0d19]{https://doi.org/10.17909/gxjc-0d19}. The transmission and UV spectra are available via \dataset[DOI: zenodo.21235410]{https://doi.org/zenodo.21235410}.}

\facility{HST (WFC3, STIS)}

\software{\texttt{astropy} \citep{astropy13,astropy18,astropy22}, \texttt{batman} \citep{batman}, \texttt{corner} \citep{corner}, \texttt{emcee} \citep{emcee}, \texttt{ExoTIC-LD} \citep{exoticld}, \texttt{lyapy} \citep{lyapy}, \texttt{matplotlib} \citep{matplotlib}, \texttt{MultiNest} \citep{multinest09,multinest13,multinest19}, \texttt{numpy} \citep{numpy}, \texttt{PACMAN} \citep{pacman}, \texttt{pandas} \citep{pandas}, \texttt{petitRADTRANS} \citep{prt1,prt2}, \texttt{PyMultiNest} \citep{pymultinest}, \texttt{scipy} \citep{scipy} }

\bibliography{literature}{}

@ARTICLE{pacman,
       author = {{Zieba}, Sebastian and {Kreidberg}, Laura},
        title = "{PACMAN: A pipeline to reduce and analyze Hubble Wide Field Camera 3 IR Grism data}",
      journal = {The Journal of Open Source Software},
         year = 2022,
        month = dec,
       volume = {7},
       number = {80},
          eid = {4838},
        pages = {4838},
          doi = {10.21105/joss.04838},
archivePrefix = {arXiv},
       eprint = {2212.11421},
 primaryClass = {astro-ph.IM},
       adsurl = {https://ui.adsabs.harvard.edu/abs/2022JOSS....7.4838Z}
}

@ARTICLE{thorngren26,
       author = {{Thorngren}, Daniel P. and {Sing}, David K. and {Mukherjee}, Sagnick},
        title = "{Bayesian Model Comparison and Significance: Widespread Errors and How to Correct Them}",
      journal = {\apjs},
         year = 2026,
        month = mar,
       volume = {283},
       number = {1},
          eid = {10},
        pages = {10},
          doi = {10.3847/1538-4365/ae0e71},
archivePrefix = {arXiv},
       eprint = {2510.00169},
 primaryClass = {astro-ph.EP},
       adsurl = {https://ui.adsabs.harvard.edu/abs/2026ApJS..283...10T}
}

@ARTICLE{kahle25,
       author = {{Kahle}, K. Angelique and {Blecic}, Jasmina and {Ashtari}, Reza and {Kreidberg}, Laura and {Kawashima}, Yui and {Cubillos}, Patricio E. and {Deming}, Drake and {Jenkins}, James S. and {Molli{\`e}re}, Paul and {Redfield}, Seth and {Tian}, Qiushi Chris and {Vines}, Jose I. and {Wilson}, David J. and {Acu{\~n}a}, Lorena and {Bitsch}, Bertram and {Brande}, Jonathan and {France}, Kevin and {Stevenson}, Kevin B. and {Crossfield}, Ian J.~M. and {Daylan}, Tansu and {Dobbs-Dixon}, Ian and {Evans-Soma}, Thomas M. and {Gapp}, Cyril and {Garc{\'\i}a Mu{\~n}oz}, Antonio and {Heng}, Kevin and {Hu}, Renyu and {Shkolnik}, Evgenya L. and {Stassun}, Keivan G. and {Teske}, Johanna},
        title = "{The SPACE Program: I. The featureless spectrum of HD 86226 c challenges sub-Neptune atmosphere trends}",
      journal = {\aap},
         year = 2025,
        month = sep,
       volume = {701},
          eid = {A184},
        pages = {A184},
          doi = {10.1051/0004-6361/202554916},
archivePrefix = {arXiv},
       eprint = {2507.13439},
 primaryClass = {astro-ph.EP},
       adsurl = {https://ui.adsabs.harvard.edu/abs/2025A&A...701A.184K}
}

@ARTICLE{deming13,
       author = {{Deming}, Drake and {Wilkins}, Ashlee and {McCullough}, Peter and {Burrows}, Adam and {Fortney}, Jonathan J. and {Agol}, Eric and {Dobbs-Dixon}, Ian and {Madhusudhan}, Nikku and {Crouzet}, Nicolas and {Desert}, Jean-Michel and {Gilliland}, Ronald L. and {Haynes}, Korey and {Knutson}, Heather A. and {Line}, Michael and {Magic}, Zazralt and {Mandell}, Avi M. and {Ranjan}, Sukrit and {Charbonneau}, David and {Clampin}, Mark and {Seager}, Sara and {Showman}, Adam P.},
        title = "{Infrared Transmission Spectroscopy of the Exoplanets HD 209458b and XO-1b Using the Wide Field Camera-3 on the Hubble Space Telescope}",
      journal = {\apj},
         year = 2013,
        month = sep,
       volume = {774},
       number = {2},
          eid = {95},
        pages = {95},
          doi = {10.1088/0004-637X/774/2/95},
archivePrefix = {arXiv},
       eprint = {1302.1141},
 primaryClass = {astro-ph.EP},
       adsurl = {https://ui.adsabs.harvard.edu/abs/2013ApJ...774...95D}
}

@ARTICLE{bachmann25,
       author = {{Bachmann}, N. and {Kreidberg}, L. and {Molli{\`e}re}, P. and {Deming}, D. and {Tsai}, S.-M.},
        title = "{Osiris revisited: Confirming a solar metallicity and low C/O in HD 209458 b}",
      journal = {\aap},
         year = 2025,
        month = aug,
       volume = {700},
          eid = {A105},
        pages = {A105},
          doi = {10.1051/0004-6361/202555577},
archivePrefix = {arXiv},
       eprint = {2506.16232},
 primaryClass = {astro-ph.EP},
       adsurl = {https://ui.adsabs.harvard.edu/abs/2025A&A...700A.105B}
}

@ARTICLE{kreidberg14,
       author = {{Kreidberg}, Laura and {Bean}, Jacob L. and {D{\'e}sert}, Jean-Michel and {Benneke}, Bj{\"o}rn and {Deming}, Drake and {Stevenson}, Kevin B. and {Seager}, Sara and {Berta-Thompson}, Zachory and {Seifahrt}, Andreas and {Homeier}, Derek},
        title = "{Clouds in the atmosphere of the super-Earth exoplanet GJ1214b}",
      journal = {\nat},
         year = 2014,
        month = jan,
       volume = {505},
       number = {7481},
        pages = {69-72},
          doi = {10.1038/nature12888},
archivePrefix = {arXiv},
       eprint = {1401.0022},
 primaryClass = {astro-ph.EP},
       adsurl = {https://ui.adsabs.harvard.edu/abs/2014Natur.505...69K}
}

@ARTICLE{kreidberg18,
       author = {{Kreidberg}, Laura and {Line}, Michael R. and {Parmentier}, Vivien and {Stevenson}, Kevin B. and {Louden}, Tom and {Bonnefoy}, Mick{\"a}el and {Faherty}, Jacqueline K. and {Henry}, Gregory W. and {Williamson}, Michael H. and {Stassun}, Keivan and {Beatty}, Thomas G. and {Bean}, Jacob L. and {Fortney}, Jonathan J. and {Showman}, Adam P. and {D{\'e}sert}, Jean-Michel and {Arcangeli}, Jacob},
        title = "{Global Climate and Atmospheric Composition of the Ultra-hot Jupiter WASP-103b from HST and Spitzer Phase Curve Observations}",
      journal = {\aj},
         year = 2018,
        month = jul,
       volume = {156},
       number = {1},
          eid = {17},
        pages = {17},
          doi = {10.3847/1538-3881/aac3df},
archivePrefix = {arXiv},
       eprint = {1805.00029},
 primaryClass = {astro-ph.EP},
       adsurl = {https://ui.adsabs.harvard.edu/abs/2018AJ....156...17K}
}

@ARTICLE{badenas_agusti20,
       author = {{Badenas-Agusti}, Mariona and {G{\"u}nther}, Maximilian N. and {Daylan}, Tansu and {Mikal-Evans}, Thomas and {Vanderburg}, Andrew and {Huang}, Chelsea X. and {Matthews}, Elisabeth and {Rackham}, Benjamin V. and {Bieryla}, Allyson and {Stassun}, Keivan G. and {Kane}, Stephen R. and {Shporer}, Avi and {Fulton}, Benjamin J. and {Hill}, Michelle L. and {Nowak}, Grzegorz and {Ribas}, Ignasi and {Pall{\'e}}, Enric and {Jenkins}, Jon M. and {Latham}, David W. and {Seager}, Sara and {Ricker}, George R. and {Vanderspek}, Roland K. and {Winn}, Joshua N. and {Abril-Pla}, Oriol and {Collins}, Karen A. and {Serra}, Pere Guerra and {Niraula}, Prajwal and {Rustamkulov}, Zafar and {Barclay}, Thomas and {Crossfield}, Ian J.~M. and {Howell}, Steve B. and {Ciardi}, David R. and {Gonzales}, Erica J. and {Schlieder}, Joshua E. and {Caldwell}, Douglas A. and {Fausnaugh}, Michael and {McDermott}, Scott and {Paegert}, Martin and {Pepper}, Joshua and {Rose}, Mark E. and {Twicken}, Joseph D.},
        title = "{HD 191939: Three Sub-Neptunes Transiting a Sun-like Star Only 54 pc Away}",
      journal = {\aj},
         year = 2020,
        month = sep,
       volume = {160},
       number = {3},
          eid = {113},
        pages = {113},
          doi = {10.3847/1538-3881/aba0b5},
archivePrefix = {arXiv},
       eprint = {2002.03958},
 primaryClass = {astro-ph.EP},
       adsurl = {https://ui.adsabs.harvard.edu/abs/2020AJ....160..113B}
}

@ARTICLE{lubin22,
       author = {{Lubin}, Jack and {Van Zandt}, Judah and {Holcomb}, Rae and {Weiss}, Lauren M. and {Petigura}, Erik A. and {Robertson}, Paul and {Akana Murphy}, Joseph M. and {Scarsdale}, Nicholas and {Batygin}, Konstantin and {Polanski}, Alex S. and {Batalha}, Natalie M. and {Crossfield}, Ian J.~M. and {Dressing}, Courtney and {Fulton}, Benjamin and {Howard}, Andrew W. and {Huber}, Daniel and {Isaacson}, Howard and {Kane}, Stephen R. and {Roy}, Arpita and {Beard}, Corey and {Blunt}, Sarah and {Chontos}, Ashley and {Dai}, Fei and {Dalba}, Paul A. and {Gary}, Kaz and {Giacalone}, Steven and {Hill}, Michelle L. and {Mayo}, Andrew and {Mo{\v{c}}nik}, Teo and {Kosiarek}, Molly R. and {Rice}, Malena and {Rubenzahl}, Ryan A. and {Latham}, David W. and {Seager}, S. and {Winn}, Joshua N. and {Gary}, Kaz},
        title = "{TESS-Keck Survey. IX. Masses of Three Sub-Neptunes Orbiting HD 191939 and the Discovery of a Warm Jovian plus a Distant Substellar Companion}",
      journal = {\aj},
         year = 2022,
        month = feb,
       volume = {163},
       number = {2},
          eid = {101},
        pages = {101},
          doi = {10.3847/1538-3881/ac3d38},
archivePrefix = {arXiv},
       eprint = {2108.02208},
 primaryClass = {astro-ph.EP},
       adsurl = {https://ui.adsabs.harvard.edu/abs/2022AJ....163..101L}
}

@ARTICLE{lubin24,
       author = {{Lubin}, Jack and {Petigura}, Erik A. and {Van Zandt}, Judah and {Beard}, Corey and {Dai}, Fei and {Halverson}, Samuel and {Holcomb}, Rae and {Howard}, Andrew W. and {Isaacson}, Howard and {Luhn}, Jacob and {Robertson}, Paul and {Rubenzahl}, Ryan A. and {Stef{\'a}nsson}, Gu{\dj}mundur and {Winn}, Joshua N. and {Brodheim}, Max and {Deich}, William and {Hill}, Grant M. and {Gibson}, Steven R. and {Holden}, Bradford and {Householder}, Aaron and {Laher}, Russ R. and {Lanclos}, Kyle and {Payne}, Joel and {Roy}, Arpita and {Smith}, Roger and {Shaum}, Abby P. and {Schwab}, Christian and {Walawender}, Josh},
        title = "{The HD 191939 Exoplanet System is Well Aligned and Flat}",
      journal = {\aj},
         year = 2024,
        month = nov,
       volume = {168},
       number = {5},
          eid = {196},
        pages = {196},
          doi = {10.3847/1538-3881/ad79ed},
archivePrefix = {arXiv},
       eprint = {2409.06795},
 primaryClass = {astro-ph.EP},
       adsurl = {https://ui.adsabs.harvard.edu/abs/2024AJ....168..196L}
}

@ARTICLE{batman,
       author = {{Kreidberg}, Laura},
        title = "{batman: BAsic Transit Model cAlculatioN in Python}",
      journal = {\pasp},
         year = 2015,
        month = nov,
       volume = {127},
       number = {957},
        pages = {1161},
          doi = {10.1086/683602},
archivePrefix = {arXiv},
       eprint = {1507.08285},
 primaryClass = {astro-ph.EP},
       adsurl = {https://ui.adsabs.harvard.edu/abs/2015PASP..127.1161K}
}

@ARTICLE{orellmiquel23,
       author = {{Orell-Miquel}, J. and {Nowak}, G. and {Murgas}, F. and {Palle}, E. and {Morello}, G. and {Luque}, R. and {Badenas-Agusti}, M. and {Ribas}, I. and {Lafarga}, M. and {Espinoza}, N. and {Morales}, J.~C. and {Zechmeister}, M. and {Alqasim}, A. and {Cochran}, W.~D. and {Gandolfi}, D. and {Goffo}, E. and {Kab{\'a}th}, P. and {Korth}, J. and {Lam}, K.~W.~F. and {Livingston}, J. and {Muresan}, A. and {Persson}, C.~M. and {Van Eylen}, V.},
        title = "{HD 191939 revisited: New and refined planet mass determinations, and a new planet in the habitable zone}",
      journal = {\aap},
         year = 2023,
        month = jan,
       volume = {669},
          eid = {A40},
        pages = {A40},
          doi = {10.1051/0004-6361/202244120},
archivePrefix = {arXiv},
       eprint = {2211.00667},
 primaryClass = {astro-ph.EP},
       adsurl = {https://ui.adsabs.harvard.edu/abs/2023A&A...669A..40O}
}

@MISC{exoticld,
       author = {{Grant}, David and {Wakeford}, Hannah R.},
        title = "{Exo-TiC/ExoTiC-LD: ExoTiC-LD v3.0.0}",
 howpublished = {Zenodo},
         year = 2022,
        month = dec,
          eid = {10.5281/zenodo.7437681},
          doi = {10.5281/zenodo.7437681},
      version = {v3.0.0},
    publisher = {Zenodo},
       adsurl = {https://ui.adsabs.harvard.edu/abs/2022zndo...7437681G}
}

@ARTICLE{staggergrid,
       author = {{Magic}, Z. and {Chiavassa}, A. and {Collet}, R. and {Asplund}, M.},
        title = "{The Stagger-grid: A grid of 3D stellar atmosphere models. IV. Limb darkening coefficients}",
      journal = {\aap},
         year = 2015,
        month = jan,
       volume = {573},
          eid = {A90},
        pages = {A90},
          doi = {10.1051/0004-6361/201423804},
archivePrefix = {arXiv},
       eprint = {1403.3487},
 primaryClass = {astro-ph.SR},
       adsurl = {https://ui.adsabs.harvard.edu/abs/2015A&A...573A..90M}
}

@ARTICLE{emcee,
       author = {{Foreman-Mackey}, Daniel and {Hogg}, David W. and {Lang}, Dustin and {Goodman}, Jonathan},
        title = "{emcee: The MCMC Hammer}",
      journal = {\pasp},
         year = 2013,
        month = mar,
       volume = {125},
       number = {925},
        pages = {306},
          doi = {10.1086/670067},
archivePrefix = {arXiv},
       eprint = {1202.3665},
 primaryClass = {astro-ph.IM},
       adsurl = {https://ui.adsabs.harvard.edu/abs/2013PASP..125..306F}
}

@ARTICLE{prt1,
       author = {{Molli{\`e}re}, P. and {Wardenier}, J.~P. and {van Boekel}, R. and {Henning}, Th. and {Molaverdikhani}, K. and {Snellen}, I.~A.~G.},
        title = "{petitRADTRANS. A Python radiative transfer package for exoplanet characterization and retrieval}",
      journal = {\aap},
         year = 2019,
        month = jul,
       volume = {627},
          eid = {A67},
        pages = {A67},
          doi = {10.1051/0004-6361/201935470},
archivePrefix = {arXiv},
       eprint = {1904.11504},
 primaryClass = {astro-ph.EP},
       adsurl = {https://ui.adsabs.harvard.edu/abs/2019A&A...627A..67M}
}

@ARTICLE{prt2,
       author = {{Nasedkin}, Evert and {Molli{\`e}re}, Paul and {Blain}, Doriann},
        title = "{Atmospheric Retrievals with petitRADTRANS}",
      journal = {The Journal of Open Source Software},
         year = 2024,
        month = apr,
       volume = {9},
       number = {96},
          eid = {5875},
        pages = {5875},
          doi = {10.21105/joss.05875},
archivePrefix = {arXiv},
       eprint = {2309.06755},
 primaryClass = {astro-ph.EP},
       adsurl = {https://ui.adsabs.harvard.edu/abs/2024JOSS....9.5875N}
}

@ARTICLE{h2o_opacities,
       author = {{Polyansky}, Oleg L. and {Kyuberis}, Aleksandra A. and {Zobov}, Nikolai F. and {Tennyson}, Jonathan and {Yurchenko}, Sergei N. and {Lodi}, Lorenzo},
        title = "{ExoMol molecular line lists XXX: a complete high-accuracy line list for water}",
      journal = {\mnras},
         year = 2018,
        month = oct,
       volume = {480},
       number = {2},
        pages = {2597-2608},
          doi = {10.1093/mnras/sty1877},
archivePrefix = {arXiv},
       eprint = {1807.04529},
 primaryClass = {astro-ph.EP},
       adsurl = {https://ui.adsabs.harvard.edu/abs/2018MNRAS.480.2597P}
}

@ARTICLE{co_opacities,
       author = {{Rothman}, L.~S. and {Gordon}, I.~E. and {Barber}, R.~J. and {Dothe}, H. and {Gamache}, R.~R. and {Goldman}, A. and {Perevalov}, V.~I. and {Tashkun}, S.~A. and {Tennyson}, J.},
        title = "{HITEMP, the high-temperature molecular spectroscopic database}",
      journal = {\jqsrt},
         year = 2010,
        month = oct,
       volume = {111},
        pages = {2139-2150},
          doi = {10.1016/j.jqsrt.2010.05.001},
       adsurl = {https://ui.adsabs.harvard.edu/abs/2010JQSRT.111.2139R}
}

@ARTICLE{ch4_opacities,
       author = {{Hargreaves}, Robert J. and {Gordon}, Iouli E. and {Rey}, Michael and {Nikitin}, Andrei V. and {Tyuterev}, Vladimir G. and {Kochanov}, Roman V. and {Rothman}, Laurence S.},
        title = "{An Accurate, Extensive, and Practical Line List of Methane for the HITEMP Database}",
      journal = {\apjs},
         year = 2020,
        month = apr,
       volume = {247},
       number = {2},
          eid = {55},
        pages = {55},
          doi = {10.3847/1538-4365/ab7a1a},
archivePrefix = {arXiv},
       eprint = {2001.05037},
 primaryClass = {astro-ph.EP},
       adsurl = {https://ui.adsabs.harvard.edu/abs/2020ApJS..247...55H}
}

@ARTICLE{co2_opacities,
       author = {{Yurchenko}, S.~N. and {Mellor}, Thomas M. and {Freedman}, Richard S. and {Tennyson}, J.},
        title = "{ExoMol line lists - XXXIX. Ro-vibrational molecular line list for CO$_{2}$}",
      journal = {\mnras},
         year = 2020,
        month = aug,
       volume = {496},
       number = {4},
        pages = {5282-5291},
          doi = {10.1093/mnras/staa1874},
archivePrefix = {arXiv},
       eprint = {2007.02122},
 primaryClass = {astro-ph.EP},
       adsurl = {https://ui.adsabs.harvard.edu/abs/2020MNRAS.496.5282Y}
}

@ARTICLE{nh3_opacities,
       author = {{Coles}, Phillip A. and {Yurchenko}, Sergei N. and {Tennyson}, Jonathan},
        title = "{ExoMol molecular line lists - XXXV. A rotation-vibration line list for hot ammonia}",
      journal = {\mnras},
         year = 2019,
        month = dec,
       volume = {490},
       number = {4},
        pages = {4638-4647},
          doi = {10.1093/mnras/stz2778},
archivePrefix = {arXiv},
       eprint = {1911.10369},
 primaryClass = {astro-ph.SR},
       adsurl = {https://ui.adsabs.harvard.edu/abs/2019MNRAS.490.4638C}
}

@ARTICLE{brande24,
       author = {{Brande}, Jonathan and {Crossfield}, Ian J.~M. and {Kreidberg}, Laura and {Morley}, Caroline V. and {Barman}, Travis and {Benneke}, Bj{\"o}rn and {Christiansen}, Jessie L. and {Dragomir}, Diana and {Fortney}, Jonathan J. and {Greene}, Thomas P. and {Hardegree-Ullman}, Kevin K. and {Howard}, Andrew W. and {Knutson}, Heather A. and {Lothringer}, Joshua D. and {Mikal-Evans}, Thomas},
        title = "{Clouds and Clarity: Revisiting Atmospheric Feature Trends in Neptune-size Exoplanets}",
      journal = {\apjl},
         year = 2024,
        month = jan,
       volume = {961},
       number = {1},
          eid = {L23},
        pages = {L23},
          doi = {10.3847/2041-8213/ad1b5c},
archivePrefix = {arXiv},
       eprint = {2310.07714},
 primaryClass = {astro-ph.EP},
       adsurl = {https://ui.adsabs.harvard.edu/abs/2024ApJ...961L..23B}
}

@software{lyapy,
       author = {{Youngblood}, Allison and {Newton}, Elisabeth R.},
        title = "{allisony/lyapy: First release created for citation purposes in the literature}",
         year = 2022,
        month = aug,
          eid = {10.5281/zenodo.6949067},
          doi = {10.5281/zenodo.6949067},
      version = {v1.0.0},
    publisher = {Zenodo},
       adsurl = {https://ui.adsabs.harvard.edu/abs/2022zndo...6949067Y}
}

@ARTICLE{woods09,
       author = {{Woods}, Thomas N. and {Chamberlin}, Phillip C. and {Harder}, Jerald W. and {Hock}, Rachel A. and {Snow}, Martin and {Eparvier}, Francis G. and {Fontenla}, Juan and {McClintock}, William E. and {Richard}, Erik C.},
        title = "{Solar Irradiance Reference Spectra (SIRS) for the 2008 Whole Heliosphere Interval (WHI)}",
      journal = {\grl},
         year = 2009,
        month = jan,
       volume = {36},
       number = {1},
          eid = {L01101},
        pages = {L01101},
          doi = {10.1029/2008GL036373},
       adsurl = {https://ui.adsabs.harvard.edu/abs/2009GeoRL..36.1101W}
}

@ARTICLE{zeng19,
       author = {{Zeng}, Li and {Jacobsen}, Stein B. and {Sasselov}, Dimitar D. and {Petaev}, Michail I. and {Vanderburg}, Andrew and {Lopez-Morales}, Mercedes and {Perez-Mercader}, Juan and {Mattsson}, Thomas R. and {Li}, Gongjie and {Heising}, Matthew Z. and {Bonomo}, Aldo S. and {Damasso}, Mario and {Berger}, Travis A. and {Cao}, Hao and {Levi}, Amit and {Wordsworth}, Robin D.},
        title = "{Growth model interpretation of planet size distribution}",
      journal = {Proceedings of the National Academy of Science},
         year = 2019,
        month = may,
       volume = {116},
       number = {20},
        pages = {9723-9728},
          doi = {10.1073/pnas.1812905116},
archivePrefix = {arXiv},
       eprint = {1906.04253},
 primaryClass = {astro-ph.EP},
       adsurl = {https://ui.adsabs.harvard.edu/abs/2019PNAS..116.9723Z}
}

@ARTICLE{aguichine21,
       author = {{Aguichine}, Artyom and {Mousis}, Olivier and {Deleuil}, Magali and {Marcq}, Emmanuel},
        title = "{Mass-Radius Relationships for Irradiated Ocean Planets}",
      journal = {\apj},
         year = 2021,
        month = jun,
       volume = {914},
       number = {2},
          eid = {84},
        pages = {84},
          doi = {10.3847/1538-4357/abfa99},
archivePrefix = {arXiv},
       eprint = {2105.01102},
 primaryClass = {astro-ph.EP},
       adsurl = {https://ui.adsabs.harvard.edu/abs/2021ApJ...914...84A}
}

@ARTICLE{morley15,
       author = {{Morley}, Caroline V. and {Fortney}, Jonathan J. and {Marley}, Mark S. and {Zahnle}, Kevin and {Line}, Michael and {Kempton}, Eliza and {Lewis}, Nikole and {Cahoy}, Kerri},
        title = "{Thermal Emission and Reflected Light Spectra of Super Earths with Flat Transmission Spectra}",
      journal = {\apj},
         year = 2015,
        month = dec,
       volume = {815},
       number = {2},
          eid = {110},
        pages = {110},
          doi = {10.1088/0004-637X/815/2/110},
archivePrefix = {arXiv},
       eprint = {1511.01492},
 primaryClass = {astro-ph.EP},
       adsurl = {https://ui.adsabs.harvard.edu/abs/2015ApJ...815..110M}
}

@ARTICLE{zahnle09,
       author = {{Zahnle}, K. and {Marley}, M.~S. and {Fortney}, J.~J.},
        title = "{Thermometric Soots on Warm Jupiters?}",
      journal = {arXiv e-prints},
         year = 2009,
        month = nov,
          eid = {arXiv:0911.0728},
        pages = {arXiv:0911.0728},
          doi = {10.48550/arXiv.0911.0728},
archivePrefix = {arXiv},
       eprint = {0911.0728},
 primaryClass = {astro-ph.EP},
       adsurl = {https://ui.adsabs.harvard.edu/abs/2009arXiv0911.0728Z}
}

@ARTICLE{ashtari25,
       author = {{Ashtari}, Reza and {Stevenson}, Kevin B. and {Sing}, David and {L{\'o}pez-Morales}, Mercedes and {Alam}, Munazza K. and {Nikolov}, Nikolay K. and {Evans-Soma}, Thomas M.},
        title = "{The Clear Sky Corridor: Insights Towards Aerosol Formation in Exoplanets Using an AI-based Survey of Exoplanet Atmospheres}",
      journal = {\aj},
         year = 2025,
        month = feb,
       volume = {169},
       number = {2},
          eid = {106},
        pages = {106},
          doi = {10.3847/1538-3881/ada353},
archivePrefix = {arXiv},
       eprint = {2410.06804},
 primaryClass = {astro-ph.EP},
       adsurl = {https://ui.adsabs.harvard.edu/abs/2025AJ....169..106A}
}

@ARTICLE{bell22,
       author = {{Bell}, Taylor and {Ahrer}, Eva-Maria and {Brande}, Jonathan and {Carter}, Aarynn and {Feinstein}, Adina and {Guzman Caloca}, Giannina and {Mansfield}, Megan and {Zieba}, Sebastian and {Piaulet}, Caroline and {Benneke}, Bj{\"o}rn and {Filippazzo}, Joseph and {May}, Erin and {Roy}, Pierre-Alexis and {Kreidberg}, Laura and {Stevenson}, Kevin},
        title = "{Eureka!: An End-to-End Pipeline for JWST Time-Series Observations}",
      journal = {The Journal of Open Source Software},
         year = 2022,
        month = nov,
       volume = {7},
       number = {79},
          eid = {4503},
        pages = {4503},
          doi = {10.21105/joss.04503},
archivePrefix = {arXiv},
       eprint = {2207.03585},
 primaryClass = {astro-ph.IM},
       adsurl = {https://ui.adsabs.harvard.edu/abs/2022JOSS....7.4503B}
}

@ARTICLE{pont06,
       author = {{Pont}, Fr{\'e}d{\'e}ric and {Zucker}, Shay and {Queloz}, Didier},
        title = "{The effect of red noise on planetary transit detection}",
      journal = {\mnras},
         year = 2006,
        month = nov,
       volume = {373},
       number = {1},
        pages = {231-242},
          doi = {10.1111/j.1365-2966.2006.11012.x},
archivePrefix = {arXiv},
       eprint = {astro-ph/0608597},
 primaryClass = {astro-ph},
       adsurl = {https://ui.adsabs.harvard.edu/abs/2006MNRAS.373..231P}
}

@book{Gregory05,
    place={Cambridge},
    title={Bayesian Logical Data Analysis for the Physical Sciences: A Comparative Approach with Mathematica® Support},
    publisher={Cambridge University Press},
    author={Gregory, Phil},
    year={2005}
}

@ARTICLE{crossfield17,
       author = {{Crossfield}, Ian J.~M. and {Kreidberg}, Laura},
        title = "{Trends in Atmospheric Properties of Neptune-size Exoplanets}",
      journal = {\aj},
         year = 2017,
        month = dec,
       volume = {154},
       number = {6},
          eid = {261},
        pages = {261},
          doi = {10.3847/1538-3881/aa9279},
archivePrefix = {arXiv},
       eprint = {1708.00016},
 primaryClass = {astro-ph.EP},
       adsurl = {https://ui.adsabs.harvard.edu/abs/2017AJ....154..261C}
}

@ARTICLE{chachan25,
       author = {{Chachan}, Yayaati and {Fortney}, Jonathan J. and {Ohno}, Kazumasa and {Thorngren}, Daniel and {Murray-Clay}, Ruth},
        title = "{Revising the Giant Planet Mass─Metallicity Relation: Deciphering the Formation Sequence of Giant Planets}",
      journal = {\apj},
         year = 2025,
        month = nov,
       volume = {994},
       number = {1},
          eid = {43},
        pages = {43},
          doi = {10.3847/1538-4357/ae0cbf},
archivePrefix = {arXiv},
       eprint = {2509.20428},
 primaryClass = {astro-ph.EP},
       adsurl = {https://ui.adsabs.harvard.edu/abs/2025ApJ...994...43C}
}

@ARTICLE{thorngren16,
       author = {{Thorngren}, Daniel P. and {Fortney}, Jonathan J. and {Murray-Clay}, Ruth A. and {Lopez}, Eric D.},
        title = "{The Mass-Metallicity Relation for Giant Planets}",
      journal = {\apj},
         year = 2016,
        month = nov,
       volume = {831},
       number = {1},
          eid = {64},
        pages = {64},
          doi = {10.3847/0004-637X/831/1/64},
archivePrefix = {arXiv},
       eprint = {1511.07854},
 primaryClass = {astro-ph.EP},
       adsurl = {https://ui.adsabs.harvard.edu/abs/2016ApJ...831...64T}
}

@ARTICLE{kempton23,
       author = {{Kempton}, Eliza M.-R. and {Zhang}, Michael and {Bean}, Jacob L. and {Steinrueck}, Maria E. and {Piette}, Anjali A.~A. and {Parmentier}, Vivien and {Malsky}, Isaac and {Roman}, Michael T. and {Rauscher}, Emily and {Gao}, Peter and {Bell}, Taylor J. and {Xue}, Qiao and {Taylor}, Jake and {Savel}, Arjun B. and {Arnold}, Kenneth E. and {Nixon}, Matthew C. and {Stevenson}, Kevin B. and {Mansfield}, Megan and {Kendrew}, Sarah and {Zieba}, Sebastian and {Ducrot}, Elsa and {Dyrek}, Achr{\`e}ne and {Lagage}, Pierre-Olivier and {Stassun}, Keivan G. and {Henry}, Gregory W. and {Barman}, Travis and {Lupu}, Roxana and {Malik}, Matej and {Kataria}, Tiffany and {Ih}, Jegug and {Fu}, Guangwei and {Welbanks}, Luis and {McGill}, Peter},
        title = "{A reflective, metal-rich atmosphere for GJ 1214b from its JWST phase curve}",
      journal = {\nat},
         year = 2023,
        month = aug,
       volume = {620},
       number = {7972},
        pages = {67-71},
          doi = {10.1038/s41586-023-06159-5},
archivePrefix = {arXiv},
       eprint = {2305.06240},
 primaryClass = {astro-ph.EP},
       adsurl = {https://ui.adsabs.harvard.edu/abs/2023Natur.620...67K}
}

@ARTICLE{roy23,
       author = {{Roy}, Pierre-Alexis and {Benneke}, Bj{\"o}rn and {Piaulet}, Caroline and {Gully-Santiago}, Michael A. and {Crossfield}, Ian J.~M. and {Morley}, Caroline V. and {Kreidberg}, Laura and {Mikal-Evans}, Thomas and {Brande}, Jonathan and {Delisle}, Simon and {Greene}, Thomas P. and {Hardegree-Ullman}, Kevin K. and {Barman}, Travis and {Christiansen}, Jessie L. and {Dragomir}, Diana and {Fortney}, Jonathan J. and {Howard}, Andrew W. and {Kosiarek}, Molly R. and {Lothringer}, Joshua D.},
        title = "{Water Absorption in the Transmission Spectrum of the Water World Candidate GJ 9827 d}",
      journal = {\apjl},
         year = 2023,
        month = sep,
       volume = {954},
       number = {2},
          eid = {L52},
        pages = {L52},
          doi = {10.3847/2041-8213/acebf0},
archivePrefix = {arXiv},
       eprint = {2309.10845},
 primaryClass = {astro-ph.EP},
       adsurl = {https://ui.adsabs.harvard.edu/abs/2023ApJ...954L..52R}
}

@ARTICLE{piauletghorayeb24,
       author = {{Piaulet-Ghorayeb}, Caroline and {Benneke}, Bj{\"o}rn and {Radica}, Michael and {Raul}, Eshan and {Coulombe}, Louis-Philippe and {Ahrer}, Eva-Maria and {Kubyshkina}, Daria and {Howard}, Ward S. and {Krissansen-Totton}, Joshua and {MacDonald}, Ryan J. and {Roy}, Pierre-Alexis and {Louca}, Amy and {Christie}, Duncan and {Fournier-Tondreau}, Marylou and {Allart}, Romain and {Miguel}, Yamila and {Schlichting}, Hilke E. and {Welbanks}, Luis and {Cadieux}, Charles and {Dorn}, Caroline and {Evans-Soma}, Thomas M. and {Fortney}, Jonathan J. and {Pierrehumbert}, Raymond and {Lafreni{\`e}re}, David and {Acu{\~n}a}, Lorena and {Komacek}, Thaddeus and {Innes}, Hamish and {Beatty}, Thomas G. and {Cloutier}, Ryan and {Doyon}, Ren{\'e} and {Gagnebin}, Anna and {Gapp}, Cyril and {Knutson}, Heather A.},
        title = "{JWST/NIRISS Reveals the Water-rich ``Steam World'' Atmosphere of GJ 9827 d}",
      journal = {\apjl},
         year = 2024,
        month = oct,
       volume = {974},
       number = {1},
          eid = {L10},
        pages = {L10},
          doi = {10.3847/2041-8213/ad6f00},
archivePrefix = {arXiv},
       eprint = {2410.03527},
 primaryClass = {astro-ph.EP},
       adsurl = {https://ui.adsabs.harvard.edu/abs/2024ApJ...974L..10P}
}

@ARTICLE{benneke24,
       author = {{Benneke}, Bj{\"o}rn and {Roy}, Pierre-Alexis and {Coulombe}, Louis-Philippe and {Radica}, Michael and {Piaulet}, Caroline and {Ahrer}, Eva-Maria and {Pierrehumbert}, Raymond and {Krissansen-Totton}, Joshua and {Schlichting}, Hilke E. and {Hu}, Renyu and {Yang}, Jeehyun and {Christie}, Duncan and {Thorngren}, Daniel and {Young}, Edward D. and {Pelletier}, Stefan and {Knutson}, Heather A. and {Miguel}, Yamila and {Evans-Soma}, Thomas M. and {Dorn}, Caroline and {Gagnebin}, Anna and {Fortney}, Jonathan J. and {Komacek}, Thaddeus and {MacDonald}, Ryan and {Raul}, Eshan and {Cloutier}, Ryan and {Acuna}, Lorena and {Lafreni{\`e}re}, David and {Cadieux}, Charles and {Doyon}, Ren{\'e} and {Welbanks}, Luis and {Allart}, Romain},
        title = "{JWST Reveals CH$_4$, CO$_2$, and H$_2$O in a Metal-rich Miscible Atmosphere on a Two-Earth-Radius Exoplanet}",
      journal = {arXiv e-prints},
         year = 2024,
        month = mar,
          eid = {arXiv:2403.03325},
        pages = {arXiv:2403.03325},
          doi = {10.48550/arXiv.2403.03325},
archivePrefix = {arXiv},
       eprint = {2403.03325},
 primaryClass = {astro-ph.EP},
       adsurl = {https://ui.adsabs.harvard.edu/abs/2024arXiv240303325B}
}

@ARTICLE{ahrer25,
       author = {{Ahrer}, Eva-Maria and {Radica}, Michael and {Piaulet-Ghorayeb}, Caroline and {Raul}, Eshan and {Wiser}, Lindsey and {Welbanks}, Luis and {Acu{\~n}a}, Lorena and {Allart}, Romain and {Coulombe}, Louis-Philippe and {Louca}, Amy and {MacDonald}, Ryan and {Saidel}, Morgan and {Evans-Soma}, Thomas M. and {Benneke}, Bj{\"o}rn and {Christie}, Duncan and {Beatty}, Thomas G. and {Cadieux}, Charles and {Cloutier}, Ryan and {Doyon}, Ren{\'e} and {Fortney}, Jonathan J. and {Gagnebin}, Anna and {Gapp}, Cyril and {Innes}, Hamish and {Knutson}, Heather A. and {Komacek}, Thaddeus and {Krissansen-Totton}, Joshua and {Miguel}, Yamila and {Pierrehumbert}, Raymond and {Roy}, Pierre-Alexis and {Schlichting}, Hilke E.},
        title = "{Escaping Helium and a Highly Muted Spectrum Suggest a Metal-enriched Atmosphere on Sub-Neptune GJ 3090 b from JWST Transit Spectroscopy}",
      journal = {\apjl},
         year = 2025,
        month = may,
       volume = {985},
       number = {1},
          eid = {L10},
        pages = {L10},
          doi = {10.3847/2041-8213/add010},
archivePrefix = {arXiv},
       eprint = {2504.20428},
 primaryClass = {astro-ph.EP},
       adsurl = {https://ui.adsabs.harvard.edu/abs/2025ApJ...985L..10A}
}

@ARTICLE{hu25,
       author = {{Hu}, Renyu and {Bello-Arufe}, Aaron and {Tokadjian}, Armen and {Yang}, Jeehyun and {Damiano}, Mario and {Roy}, Pierre-Alexis and {Coulombe}, Louis-Philippe and {Madhusudhan}, Nikku and {Constantinou}, Savvas and {Benneke}, Bj{\"o}rn},
        title = "{A water-rich interior in the temperate sub-Neptune K2-18 b revealed by JWST}",
      journal = {arXiv e-prints},
         year = 2025,
        month = jul,
          eid = {arXiv:2507.12622},
        pages = {arXiv:2507.12622},
          doi = {10.48550/arXiv.2507.12622},
archivePrefix = {arXiv},
       eprint = {2507.12622},
 primaryClass = {astro-ph.EP},
       adsurl = {https://ui.adsabs.harvard.edu/abs/2025arXiv250712622H}
}

@ARTICLE{tsiaras2016a,
       author = {{Tsiaras}, A. and {Waldmann}, I.~P. and {Rocchetto}, M. and {Varley}, R. and {Morello}, G. and {Damiano}, M. and {Tinetti}, G.},
        title = "{A New Approach to Analyzing HST Spatial Scans: The Transmission Spectrum of HD 209458 b}",
      journal = {\apj},
         year = 2016,
        month = dec,
       volume = {832},
       number = {2},
          eid = {202},
        pages = {202},
          doi = {10.3847/0004-637X/832/2/202},
archivePrefix = {arXiv},
       eprint = {1511.07796},
 primaryClass = {astro-ph.EP},
       adsurl = {https://ui.adsabs.harvard.edu/abs/2016ApJ...832..202T}
}

@ARTICLE{tsiaras2016b,
       author = {{Tsiaras}, A. and {Rocchetto}, M. and {Waldmann}, I.~P. and {Venot}, O. and {Varley}, R. and {Morello}, G. and {Damiano}, M. and {Tinetti}, G. and {Barton}, E.~J. and {Yurchenko}, S.~N. and {Tennyson}, J.},
        title = "{Detection of an Atmosphere Around the Super-Earth 55 Cancri e}",
      journal = {\apj},
         year = 2016,
        month = apr,
       volume = {820},
       number = {2},
          eid = {99},
        pages = {99},
          doi = {10.3847/0004-637X/820/2/99},
archivePrefix = {arXiv},
       eprint = {1511.08901},
 primaryClass = {astro-ph.EP},
       adsurl = {https://ui.adsabs.harvard.edu/abs/2016ApJ...820...99T}
}

@ARTICLE{tsiaras2018,
       author = {{Tsiaras}, A. and {Waldmann}, I.~P. and {Zingales}, T. and {Rocchetto}, M. and {Morello}, G. and {Damiano}, M. and {Karpouzas}, K. and {Tinetti}, G. and {McKemmish}, L.~K. and {Tennyson}, J. and {Yurchenko}, S.~N.},
        title = "{A Population Study of Gaseous Exoplanets}",
      journal = {\aj},
         year = 2018,
        month = apr,
       volume = {155},
       number = {4},
          eid = {156},
        pages = {156},
          doi = {10.3847/1538-3881/aaaf75},
archivePrefix = {arXiv},
       eprint = {1704.05413},
 primaryClass = {astro-ph.EP},
       adsurl = {https://ui.adsabs.harvard.edu/abs/2018AJ....155..156T}
}

@ARTICLE{damiano2017,
       author = {{Damiano}, M. and {Morello}, G. and {Tsiaras}, A. and {Zingales}, T. and {Tinetti}, G.},
        title = "{Near-IR Transmission Spectrum of HAT-P-32b using HST/WFC3}",
      journal = {\aj},
         year = 2017,
        month = jul,
       volume = {154},
       number = {1},
          eid = {39},
        pages = {39},
          doi = {10.3847/1538-3881/aa738b},
archivePrefix = {arXiv},
       eprint = {1802.10010},
 primaryClass = {astro-ph.EP},
       adsurl = {https://ui.adsabs.harvard.edu/abs/2017AJ....154...39D}
}

@ARTICLE{claret2000,
       author = {{Claret}, A.},
        title = "{A new non-linear limb-darkening law for LTE stellar atmosphere models. Calculations for -5.0 <= log[M/H] <= +1, 2000 K <= T$_{eff}$ <= 50000 K at several surface gravities}",
      journal = {\aap},
         year = 2000,
        month = nov,
       volume = {363},
        pages = {1081-1190},
       adsurl = {https://ui.adsabs.harvard.edu/abs/2000A&A...363.1081C}
}

@ARTICLE{husser2013,
       author = {{Husser}, T.-O. and {Wende-von Berg}, S. and {Dreizler}, S. and {Homeier}, D. and {Reiners}, A. and {Barman}, T. and {Hauschildt}, P.~H.},
        title = "{A new extensive library of PHOENIX stellar atmospheres and synthetic spectra}",
      journal = {\aap},
         year = 2013,
        month = may,
       volume = {553},
          eid = {A6},
        pages = {A6},
          doi = {10.1051/0004-6361/201219058},
archivePrefix = {arXiv},
       eprint = {1303.5632},
 primaryClass = {astro-ph.SR},
       adsurl = {https://ui.adsabs.harvard.edu/abs/2013A&A...553A...6H}
}

@ARTICLE{fulton17,
       author = {{Fulton}, Benjamin J. and {Petigura}, Erik A. and {Howard}, Andrew W. and {Isaacson}, Howard and {Marcy}, Geoffrey W. and {Cargile}, Phillip A. and {Hebb}, Leslie and {Weiss}, Lauren M. and {Johnson}, John Asher and {Morton}, Timothy D. and {Sinukoff}, Evan and {Crossfield}, Ian J.~M. and {Hirsch}, Lea A.},
        title = "{The California-Kepler Survey. III. A Gap in the Radius Distribution of Small Planets}",
      journal = {\aj},
         year = 2017,
        month = sep,
       volume = {154},
       number = {3},
          eid = {109},
        pages = {109},
          doi = {10.3847/1538-3881/aa80eb},
archivePrefix = {arXiv},
       eprint = {1703.10375},
 primaryClass = {astro-ph.EP},
       adsurl = {https://ui.adsabs.harvard.edu/abs/2017AJ....154..109F}
}

@ARTICLE{oberg11,
       author = {{{\"O}berg}, Karin I. and {Murray-Clay}, Ruth and {Bergin}, Edwin A.},
        title = "{The Effects of Snowlines on C/O in Planetary Atmospheres}",
      journal = {\apjl},
         year = 2011,
        month = dec,
       volume = {743},
       number = {1},
          eid = {L16},
        pages = {L16},
          doi = {10.1088/2041-8205/743/1/L16},
archivePrefix = {arXiv},
       eprint = {1110.5567},
 primaryClass = {astro-ph.GA},
       adsurl = {https://ui.adsabs.harvard.edu/abs/2011ApJ...743L..16O}
}

@ARTICLE{lothringer21,
       author = {{Lothringer}, Joshua D. and {Rustamkulov}, Zafar and {Sing}, David K. and {Gibson}, Neale P. and {Wilson}, Jamie and {Schlaufman}, Kevin C.},
        title = "{A New Window into Planet Formation and Migration: Refractory-to-Volatile Elemental Ratios in Ultra-hot Jupiters}",
      journal = {\apj},
         year = 2021,
        month = jun,
       volume = {914},
       number = {1},
          eid = {12},
        pages = {12},
          doi = {10.3847/1538-4357/abf8a9},
archivePrefix = {arXiv},
       eprint = {2011.10626},
 primaryClass = {astro-ph.EP},
       adsurl = {https://ui.adsabs.harvard.edu/abs/2021ApJ...914...12L}
}

@ARTICLE{rogers10,
       author = {{Rogers}, L.~A. and {Seager}, S.},
        title = "{A Framework for Quantifying the Degeneracies of Exoplanet Interior Compositions}",
      journal = {\apj},
         year = 2010,
        month = apr,
       volume = {712},
       number = {2},
        pages = {974-991},
          doi = {10.1088/0004-637X/712/2/974},
archivePrefix = {arXiv},
       eprint = {0912.3288},
 primaryClass = {astro-ph.EP},
       adsurl = {https://ui.adsabs.harvard.edu/abs/2010ApJ...712..974R}
}

@ARTICLE{werlen25,
       author = {{Werlen}, Aaron and {Dorn}, Caroline and {Schlichting}, Hilke E. and {Grimm}, Simon L. and {Young}, Edward D.},
        title = "{Atmospheric C/O Ratios of Sub-Neptunes with Magma Oceans: Homemade rather than Inherited}",
      journal = {\apjl},
         year = 2025,
        month = aug,
       volume = {988},
       number = {2},
          eid = {L55},
        pages = {L55},
          doi = {10.3847/2041-8213/adf185},
archivePrefix = {arXiv},
       eprint = {2504.20450},
 primaryClass = {astro-ph.EP},
       adsurl = {https://ui.adsabs.harvard.edu/abs/2025ApJ...988L..55W}
}

@ARTICLE{owen13,
       author = {{Owen}, James E. and {Wu}, Yanqin},
        title = "{Kepler Planets: A Tale of Evaporation}",
      journal = {\apj},
         year = 2013,
        month = oct,
       volume = {775},
       number = {2},
          eid = {105},
        pages = {105},
          doi = {10.1088/0004-637X/775/2/105},
archivePrefix = {arXiv},
       eprint = {1303.3899},
 primaryClass = {astro-ph.EP},
       adsurl = {https://ui.adsabs.harvard.edu/abs/2013ApJ...775..105O}
}

@ARTICLE{breza25,
       author = {{Breza}, Bodie and {Nixon}, Matthew C. and {Kempton}, Eliza M.-R.},
        title = "{Not All Sub-Neptune Exoplanets Have Magma Oceans}",
      journal = {\apjl},
         year = 2025,
        month = nov,
       volume = {993},
       number = {2},
          eid = {L46},
        pages = {L46},
          doi = {10.3847/2041-8213/ae0c07},
archivePrefix = {arXiv},
       eprint = {2509.20429},
 primaryClass = {astro-ph.EP},
       adsurl = {https://ui.adsabs.harvard.edu/abs/2025ApJ...993L..46B}
}

@ARTICLE{exoplanetarchive,
       author = {{Christiansen}, Jessie L. and {McElroy}, Douglas L. and {Harbut}, Marcy and {Ciardi}, David R. and {Crane}, Megan and {Good}, John and {Hardegree-Ullman}, Kevin K. and {Kesseli}, Aurora Y. and {Lund}, Michael B. and {Lynn}, Meca and {Muthiar}, Ananda and {Nilsson}, Ricky and {Oluyide}, Toba and {Papin}, Michael and {Rivera}, Amalia and {Swain}, Melanie and {Susemiehl}, Nicholas D. and {Tam}, Raymond and {van Eyken}, Julian and {Beichman}, Charles},
        title = "{The NASA Exoplanet Archive and Exoplanet Follow-up Observing Program: Data, Tools, and Usage}",
      journal = {PSJ},
         year = 2025,
        month = aug,
       volume = {6},
       number = {8},
          eid = {186},
        pages = {186},
          doi = {10.3847/PSJ/ade3c2},
archivePrefix = {arXiv},
       eprint = {2506.03299},
 primaryClass = {astro-ph.EP},
       adsurl = {https://ui.adsabs.harvard.edu/abs/2025PSJ.....6..186C}
}

@ARTICLE{benneke13,
       author = {{Benneke}, Bj{\"o}rn and {Seager}, Sara},
        title = "{How to Distinguish between Cloudy Mini-Neptunes and Water/Volatile-dominated Super-Earths}",
      journal = {\apj},
         year = 2013,
        month = dec,
       volume = {778},
       number = {2},
          eid = {153},
        pages = {153},
          doi = {10.1088/0004-637X/778/2/153},
archivePrefix = {arXiv},
       eprint = {1306.6325},
 primaryClass = {astro-ph.EP},
       adsurl = {https://ui.adsabs.harvard.edu/abs/2013ApJ...778..153B}
}

@ARTICLE{frohlich13,
       author = {{Fr{\"o}hlich}, Claus},
        title = "{Total Solar Irradiance: What Have We Learned from the Last Three Cycles and the Recent Minimum?}",
      journal = {\ssr},
         year = 2013,
        month = jun,
       volume = {176},
       number = {1-4},
        pages = {237-252},
          doi = {10.1007/s11214-011-9780-1},
       adsurl = {https://ui.adsabs.harvard.edu/abs/2013SSRv..176..237F}
}

@ARTICLE{helling23,
       author = {{Helling}, Christiane and {Samra}, Dominic and {Lewis}, David and {Calder}, Robb and {Hirst}, Georgina and {Woitke}, Peter and {Baeyens}, Robin and {Carone}, Ludmila and {Herbort}, Oliver and {Chubb}, Katy L.},
        title = "{Exoplanet weather and climate regimes with clouds and thermal ionospheres. A model grid study in support of large-scale observational campaigns}",
      journal = {\aap},
         year = 2023,
        month = mar,
       volume = {671},
          eid = {A122},
        pages = {A122},
          doi = {10.1051/0004-6361/202243956},
archivePrefix = {arXiv},
       eprint = {2208.05562},
 primaryClass = {astro-ph.EP},
       adsurl = {https://ui.adsabs.harvard.edu/abs/2023A&A...671A.122H}
}

@ARTICLE{carone25,
       author = {{Carone}, Ludmila and {Helling}, Christiane and {Gernjak}, Sebastian and {Leitner}, Hanna and {Janz}, Tamara},
        title = "{Exoplanet climate characterization with transit asymmetries -- A comprehensive population study from the optical to the infrared}",
      journal = {arXiv e-prints},
         year = 2025,
        month = nov,
          eid = {arXiv:2511.01548},
        pages = {arXiv:2511.01548},
          doi = {10.48550/arXiv.2511.01548},
archivePrefix = {arXiv},
       eprint = {2511.01548},
 primaryClass = {astro-ph.EP},
       adsurl = {https://ui.adsabs.harvard.edu/abs/2025arXiv251101548C}
}

@ARTICLE{owen17,
       author = {{Owen}, James E. and {Wu}, Yanqin},
        title = "{The Evaporation Valley in the Kepler Planets}",
      journal = {\apj},
         year = 2017,
        month = sep,
       volume = {847},
       number = {1},
          eid = {29},
        pages = {29},
          doi = {10.3847/1538-4357/aa890a},
archivePrefix = {arXiv},
       eprint = {1705.10810},
 primaryClass = {astro-ph.EP},
       adsurl = {https://ui.adsabs.harvard.edu/abs/2017ApJ...847...29O}
}

@ARTICLE{luque22,
       author = {{Luque}, Rafael and {Pall{\'e}}, Enric},
        title = "{Density, not radius, separates rocky and water-rich small planets orbiting M dwarf stars}",
      journal = {Science},
         year = 2022,
        month = sep,
       volume = {377},
       number = {6611},
        pages = {1211-1214},
          doi = {10.1126/science.abl7164},
archivePrefix = {arXiv},
       eprint = {2209.03871},
 primaryClass = {astro-ph.EP},
       adsurl = {https://ui.adsabs.harvard.edu/abs/2022Sci...377.1211L}
}

@ARTICLE{venturini20,
       author = {{Venturini}, Julia and {Guilera}, Octavio M. and {Haldemann}, Jonas and {Ronco}, Mar{\'\i}a P. and {Mordasini}, Christoph},
        title = "{The nature of the radius valley. Hints from formation and evolution models}",
      journal = {\aap},
         year = 2020,
        month = nov,
       volume = {643},
          eid = {L1},
        pages = {L1},
          doi = {10.1051/0004-6361/202039141},
archivePrefix = {arXiv},
       eprint = {2008.05513},
 primaryClass = {astro-ph.EP},
       adsurl = {https://ui.adsabs.harvard.edu/abs/2020A&A...643L...1V}
}

@ARTICLE{izidoro22,
       author = {{Izidoro}, Andr{\'e} and {Schlichting}, Hilke E. and {Isella}, Andrea and {Dasgupta}, Rajdeep and {Zimmermann}, Christian and {Bitsch}, Bertram},
        title = "{The Exoplanet Radius Valley from Gas-driven Planet Migration and Breaking of Resonant Chains}",
      journal = {\apjl},
         year = 2022,
        month = nov,
       volume = {939},
       number = {2},
          eid = {L19},
        pages = {L19},
          doi = {10.3847/2041-8213/ac990d},
archivePrefix = {arXiv},
       eprint = {2210.05595},
 primaryClass = {astro-ph.EP},
       adsurl = {https://ui.adsabs.harvard.edu/abs/2022ApJ...939L..19I}
}

@ARTICLE{burn24,
       author = {{Burn}, Remo and {Mordasini}, Christoph and {Mishra}, Lokesh and {Haldemann}, Jonas and {Venturini}, Julia and {Emsenhuber}, Alexandre and {Henning}, Thomas},
        title = "{A radius valley between migrated steam worlds and evaporated rocky cores}",
      journal = {Nature Astronomy},
         year = 2024,
        month = apr,
       volume = {8},
        pages = {463-471},
          doi = {10.1038/s41550-023-02183-7},
archivePrefix = {arXiv},
       eprint = {2401.04380},
 primaryClass = {astro-ph.EP},
       adsurl = {https://ui.adsabs.harvard.edu/abs/2024NatAs...8..463B}
}

@ARTICLE{misener23,
       author = {{Misener}, William and {Schlichting}, Hilke E. and {Young}, Edward D.},
        title = "{Atmospheres as windows into sub-Neptune interiors: coupled chemistry and structure of hydrogen-silane-water envelopes}",
      journal = {\mnras},
         year = 2023,
        month = sep,
       volume = {524},
       number = {1},
        pages = {981-992},
          doi = {10.1093/mnras/stad1910},
archivePrefix = {arXiv},
       eprint = {2303.09653},
 primaryClass = {astro-ph.EP},
       adsurl = {https://ui.adsabs.harvard.edu/abs/2023MNRAS.524..981M}
}

@ARTICLE{shorttle24,
       author = {{Shorttle}, Oliver and {Jordan}, Sean and {Nicholls}, Harrison and {Lichtenberg}, Tim and {Bower}, Dan J.},
        title = "{Distinguishing Oceans of Water from Magma on Mini-Neptune K2-18b}",
      journal = {\apjl},
         year = 2024,
        month = feb,
       volume = {962},
       number = {1},
          eid = {L8},
        pages = {L8},
          doi = {10.3847/2041-8213/ad206e},
archivePrefix = {arXiv},
       eprint = {2401.05864},
 primaryClass = {astro-ph.EP},
       adsurl = {https://ui.adsabs.harvard.edu/abs/2024ApJ...962L...8S}
}

@ARTICLE{ito25,
       author = {{Ito}, Yuichi and {Kimura}, Tadahiro and {Ohno}, Kazumasa and {Fujii}, Yuka and {Ikoma}, Masahiro},
        title = "{Monosilane Worlds: Sub-Neptunes with Atmospheres Shaped by Reduced Magma Oceans}",
      journal = {\apj},
         year = 2025,
        month = jul,
       volume = {987},
       number = {2},
          eid = {174},
        pages = {174},
          doi = {10.3847/1538-4357/add3fe},
archivePrefix = {arXiv},
       eprint = {2505.03200},
 primaryClass = {astro-ph.EP},
       adsurl = {https://ui.adsabs.harvard.edu/abs/2025ApJ...987..174I}
}

@ARTICLE{heng25,
       author = {{Heng}, Kevin and {Owen}, James E. and {Tian}, Meng},
        title = "{The Gradient of Mean Molecular Weight across the Radius Valley}",
      journal = {\apj},
         year = 2025,
        month = nov,
       volume = {994},
       number = {1},
          eid = {28},
        pages = {28},
          doi = {10.3847/1538-4357/ae0acc},
archivePrefix = {arXiv},
       eprint = {2504.02499},
 primaryClass = {astro-ph.EP},
       adsurl = {https://ui.adsabs.harvard.edu/abs/2025ApJ...994...28H}
}

@ARTICLE{lammer03,
       author = {{Lammer}, H. and {Selsis}, F. and {Ribas}, I. and {Guinan}, E.~F. and {Bauer}, S.~J. and {Weiss}, W.~W.},
        title = "{Atmospheric Loss of Exoplanets Resulting from Stellar X-Ray and Extreme-Ultraviolet Heating}",
      journal = {\apjl},
         year = 2003,
        month = dec,
       volume = {598},
       number = {2},
        pages = {L121-L124},
          doi = {10.1086/380815},
       adsurl = {https://ui.adsabs.harvard.edu/abs/2003ApJ...598L.121L}
}

@ARTICLE{fortney13,
       author = {{Fortney}, Jonathan J. and {Mordasini}, Christoph and {Nettelmann}, Nadine and {Kempton}, Eliza M.-R. and {Greene}, Thomas P. and {Zahnle}, Kevin},
        title = "{A Framework for Characterizing the Atmospheres of Low-mass Low-density Transiting Planets}",
      journal = {\apj},
         year = 2013,
        month = sep,
       volume = {775},
       number = {1},
          eid = {80},
        pages = {80},
          doi = {10.1088/0004-637X/775/1/80},
archivePrefix = {arXiv},
       eprint = {1306.4329},
 primaryClass = {astro-ph.EP},
       adsurl = {https://ui.adsabs.harvard.edu/abs/2013ApJ...775...80F}
}

@ARTICLE{borucki11,
       author = {{Borucki}, William J. and {Koch}, David G. and {Basri}, Gibor and {Batalha}, Natalie and {Brown}, Timothy M. and {Bryson}, Stephen T. and {Caldwell}, Douglas and {Christensen-Dalsgaard}, J{\o}rgen and {Cochran}, William D. and {DeVore}, Edna and {Dunham}, Edward W. and {Gautier}, III, Thomas N. and {Geary}, John C. and {Gilliland}, Ronald and {Gould}, Alan and {Howell}, Steve B. and {Jenkins}, Jon M. and {Latham}, David W. and {Lissauer}, Jack J. and {Marcy}, Geoffrey W. and {Rowe}, Jason and {Sasselov}, Dimitar and {Boss}, Alan and {Charbonneau}, David and {Ciardi}, David and {Doyle}, Laurance and {Dupree}, Andrea K. and {Ford}, Eric B. and {Fortney}, Jonathan and {Holman}, Matthew J. and {Seager}, Sara and {Steffen}, Jason H. and {Tarter}, Jill and {Welsh}, William F. and {Allen}, Christopher and {Buchhave}, Lars A. and {Christiansen}, Jessie L. and {Clarke}, Bruce D. and {Das}, Santanu and {D{\'e}sert}, Jean-Michel and {Endl}, Michael and {Fabrycky}, Daniel and {Fressin}, Francois and {Haas}, Michael and {Horch}, Elliott and {Howard}, Andrew and {Isaacson}, Howard and {Kjeldsen}, Hans and {Kolodziejczak}, Jeffery and {Kulesa}, Craig and {Li}, Jie and {Lucas}, Philip W. and {Machalek}, Pavel and {McCarthy}, Donald and {MacQueen}, Phillip and {Meibom}, S{\o}ren and {Miquel}, Thibaut and {Prsa}, Andrej and {Quinn}, Samuel N. and {Quintana}, Elisa V. and {Ragozzine}, Darin and {Sherry}, William and {Shporer}, Avi and {Tenenbaum}, Peter and {Torres}, Guillermo and {Twicken}, Joseph D. and {Van Cleve}, Jeffrey and {Walkowicz}, Lucianne and {Witteborn}, Fred C. and {Still}, Martin},
        title = "{Characteristics of Planetary Candidates Observed by Kepler. II. Analysis of the First Four Months of Data}",
      journal = {\apj},
         year = 2011,
        month = jul,
       volume = {736},
       number = {1},
          eid = {19},
        pages = {19},
          doi = {10.1088/0004-637X/736/1/19},
archivePrefix = {arXiv},
       eprint = {1102.0541},
 primaryClass = {astro-ph.EP},
       adsurl = {https://ui.adsabs.harvard.edu/abs/2011ApJ...736...19B}
}

@ARTICLE{batalha13,
       author = {{Batalha}, Natalie M. and {Rowe}, Jason F. and {Bryson}, Stephen T. and {Barclay}, Thomas and {Burke}, Christopher J. and {Caldwell}, Douglas A. and {Christiansen}, Jessie L. and {Mullally}, Fergal and {Thompson}, Susan E. and {Brown}, Timothy M. and {Dupree}, Andrea K. and {Fabrycky}, Daniel C. and {Ford}, Eric B. and {Fortney}, Jonathan J. and {Gilliland}, Ronald L. and {Isaacson}, Howard and {Latham}, David W. and {Marcy}, Geoffrey W. and {Quinn}, Samuel N. and {Ragozzine}, Darin and {Shporer}, Avi and {Borucki}, William J. and {Ciardi}, David R. and {Gautier}, III, Thomas N. and {Haas}, Michael R. and {Jenkins}, Jon M. and {Koch}, David G. and {Lissauer}, Jack J. and {Rapin}, William and {Basri}, Gibor S. and {Boss}, Alan P. and {Buchhave}, Lars A. and {Carter}, Joshua A. and {Charbonneau}, David and {Christensen-Dalsgaard}, Joergen and {Clarke}, Bruce D. and {Cochran}, William D. and {Demory}, Brice-Olivier and {Desert}, Jean-Michel and {Devore}, Edna and {Doyle}, Laurance R. and {Esquerdo}, Gilbert A. and {Everett}, Mark and {Fressin}, Francois and {Geary}, John C. and {Girouard}, Forrest R. and {Gould}, Alan and {Hall}, Jennifer R. and {Holman}, Matthew J. and {Howard}, Andrew W. and {Howell}, Steve B. and {Ibrahim}, Khadeejah A. and {Kinemuchi}, Karen and {Kjeldsen}, Hans and {Klaus}, Todd C. and {Li}, Jie and {Lucas}, Philip W. and {Meibom}, S{\o}ren and {Morris}, Robert L. and {Pr{\v{s}}a}, Andrej and {Quintana}, Elisa and {Sanderfer}, Dwight T. and {Sasselov}, Dimitar and {Seader}, Shawn E. and {Smith}, Jeffrey C. and {Steffen}, Jason H. and {Still}, Martin and {Stumpe}, Martin C. and {Tarter}, Jill C. and {Tenenbaum}, Peter and {Torres}, Guillermo and {Twicken}, Joseph D. and {Uddin}, Kamal and {Van Cleve}, Jeffrey and {Walkowicz}, Lucianne and {Welsh}, William F.},
        title = "{Planetary Candidates Observed by Kepler. III. Analysis of the First 16 Months of Data}",
      journal = {\apjs},
         year = 2013,
        month = feb,
       volume = {204},
       number = {2},
          eid = {24},
        pages = {24},
          doi = {10.1088/0067-0049/204/2/24},
archivePrefix = {arXiv},
       eprint = {1202.5852},
 primaryClass = {astro-ph.EP},
       adsurl = {https://ui.adsabs.harvard.edu/abs/2013ApJS..204...24B}
}

@ARTICLE{yung84,
       author = {{Yung}, Y.~L. and {Allen}, M. and {Pinto}, J.~P.},
        title = "{Photochemistry of the atmosphere of Titan - Comparison between model and observations}",
      journal = {\apjs},
         year = 1984,
        month = jul,
       volume = {55},
        pages = {465-506},
          doi = {10.1086/190963},
       adsurl = {https://ui.adsabs.harvard.edu/abs/1984ApJS...55..465Y}
}

@ARTICLE{rackham19,
       author = {{Rackham}, Benjamin V. and {Apai}, D{\'a}niel and {Giampapa}, Mark S.},
        title = "{The Transit Light Source Effect. II. The Impact of Stellar Heterogeneity on Transmission Spectra of Planets Orbiting Broadly Sun-like Stars}",
      journal = {\aj},
         year = 2019,
        month = mar,
       volume = {157},
       number = {3},
          eid = {96},
        pages = {96},
          doi = {10.3847/1538-3881/aaf892},
archivePrefix = {arXiv},
       eprint = {1812.06184},
 primaryClass = {astro-ph.EP},
       adsurl = {https://ui.adsabs.harvard.edu/abs/2019AJ....157...96R}
}

@ARTICLE{hatp11,
       author = {{Basilicata}, M. and {Giacobbe}, P. and {Bonomo}, A.~S. and {Scandariato}, G. and {Brogi}, M. and {Singh}, V. and {Di Paola}, A. and {Mancini}, L. and {Sozzetti}, A. and {Lanza}, A.~F. and {Cubillos}, P.~E. and {Damasso}, M. and {Desidera}, S. and {Biazzo}, K. and {Bignamini}, A. and {Borsa}, F. and {Cabona}, L. and {Carleo}, I. and {Ghedina}, A. and {Guilluy}, G. and {Maggio}, A. and {Mainella}, G. and {Micela}, G. and {Molinari}, E. and {Molinaro}, M. and {Nardiello}, D. and {Pedani}, M. and {Pino}, L. and {Poretti}, E. and {Southworth}, J. and {Stangret}, M. and {Turrini}, D.},
        title = "{The GAPS Programme at TNG. LV. Multiple molecular species in the atmosphere of HAT-P-11 b and review of the HAT-P-11 planetary system}",
      journal = {\aap},
         year = 2024,
        month = jun,
       volume = {686},
          eid = {A127},
        pages = {A127},
          doi = {10.1051/0004-6361/202347659},
archivePrefix = {arXiv},
       eprint = {2403.01527},
 primaryClass = {astro-ph.EP},
       adsurl = {https://ui.adsabs.harvard.edu/abs/2024A&A...686A.127B}
}

@ARTICLE{hd106315,
       author = {{Howard}, Andrew W. and {Sinukoff}, Evan and {Blunt}, Sarah and {Petigura}, Erik A. and {Crossfield}, Ian J.~M. and {Isaacson}, Howard and {Kosiarek}, Molly and {Rubenzahl}, Ryan A. and {Brewer}, John M. and {Fulton}, Benjamin J. and {Dressing}, Courtney D. and {Hirsch}, Lea A. and {Knutson}, Heather and {Livingston}, John H. and {Mills}, Sean M. and {Roy}, Arpita and {Weiss}, Lauren M. and {Benneke}, Bjorn and {Ciardi}, David R. and {Christiansen}, Jessie L. and {Cochran}, William D. and {Crepp}, Justin R. and {Gonzales}, Erica and {Hansen}, Brad M.~S. and {Hardegree-Ullman}, Kevin and {Howell}, Steve B. and {L{\'e}pine}, S{\'e}bastien and {Martinez}, Arturo O. and {Rogers}, Leslie A. and {Schlieder}, Joshua E. and {Werner}, Michael and {Polanski}, Alex S. and {Angelo}, Isabel and {Beard}, Corey and {Behmard}, Aida and {Bouma}, Luke G. and {Brinkman}, Casey L. and {Chontos}, Ashley and {Dai}, Fei and {Dalba}, Paul A. and {Giacalone}, Steven and {Grunblatt}, Samuel K. and {Hill}, Michelle L. and {Kane}, Stephen R. and {Lubin}, Jack and {Mayo}, Andrew W. and {Mocnik}, Teo and {Murphy}, Joseph M. Akana and {Rice}, Malena and {Rosenthal}, Lee J. and {Tyler}, Dakotah and {Van Zandt}, Judah and {Yee}, Samuel W.},
        title = "{Planet Masses, Radii, and Orbits from NASA's K2 Mission}",
      journal = {\apjs},
         year = 2025,
        month = jun,
       volume = {278},
       number = {2},
          eid = {52},
        pages = {52},
          doi = {10.3847/1538-4365/adc5e4},
archivePrefix = {arXiv},
       eprint = {2502.04436},
 primaryClass = {astro-ph.EP},
       adsurl = {https://ui.adsabs.harvard.edu/abs/2025ApJS..278...52H}
}

@ARTICLE{hatp26,
       author = {{Mancini}, L. and {Esposito}, M. and {Covino}, E. and {Southworth}, J. and {Poretti}, E. and {Andreuzzi}, G. and {Barbato}, D. and {Biazzo}, K. and {Borsato}, L. and {Bruni}, I. and {Damasso}, M. and {Di Fabrizio}, L. and {Evans}, D.~F. and {Granata}, V. and {Lanza}, A.~F. and {Naponiello}, L. and {Nascimbeni}, V. and {Pinamonti}, M. and {Sozzetti}, A. and {Tregloan-Reed}, J. and {Basilicata}, M. and {Bignamini}, A. and {Bonomo}, A.~S. and {Claudi}, R. and {Cosentino}, R. and {Desidera}, S. and {Fiorenzano}, A.~F.~M. and {Giacobbe}, P. and {Harutyunyan}, A. and {Henning}, Th. and {Knapic}, C. and {Maggio}, A. and {Micela}, G. and {Molinari}, E. and {Pagano}, I. and {Pedani}, M. and {Piotto}, G.},
        title = "{The GAPS Programme at TNG. XXXVI. Measurement of the Rossiter-McLaughlin effect and revising the physical and orbital parameters of the HAT-P-15, HAT-P-17, HAT-P-21, HAT-P-26, HAT-P-29 eccentric planetary systems}",
      journal = {\aap},
         year = 2022,
        month = aug,
       volume = {664},
          eid = {A162},
        pages = {A162},
          doi = {10.1051/0004-6361/202243742},
archivePrefix = {arXiv},
       eprint = {2205.10549},
 primaryClass = {astro-ph.EP},
       adsurl = {https://ui.adsabs.harvard.edu/abs/2022A&A...664A.162M}
}

@ARTICLE{roy25,
       author = {{Roy}, Pierre-Alexis and {Benneke}, Bj{\"o}rn and {Fournier-Tondreau}, Marylou and {Coulombe}, Louis-Philippe and {Piaulet-Ghorayeb}, Caroline and {Lafreni{\`e}re}, David and {Allart}, Romain and {Cowan}, Nicolas B. and {Dang}, Lisa and {Johnstone}, Doug and {Langeveld}, Adam B. and {Pelletier}, Stefan and {Radica}, Michael and {Taylor}, Jake and {Albert}, Lo{\"\i}c and {Doyon}, Ren{\'e} and {Flagg}, Laura and {Jayawardhana}, Ray and {MacDonald}, Ryan J. and {Turner}, Jake D.},
        title = "{Diversity in the haziness and chemistry of temperate sub-Neptunes}",
      journal = {Nature Astronomy},
         year = 2025,
        month = dec,
          doi = {10.1038/s41550-025-02723-3},
archivePrefix = {arXiv},
       eprint = {2512.10876},
 primaryClass = {astro-ph.EP},
       adsurl = {https://ui.adsabs.harvard.edu/abs/2025NatAs.tmp..256R}
}

@ARTICLE{Booth2017,
       author = {{Booth}, Richard A. and {Clarke}, Cathie J. and {Madhusudhan}, Nikku and {Ilee}, John D.},
        title = "{Chemical enrichment of giant planets and discs due to pebble drift}",
      journal = {\mnras},
         year = 2017,
        month = aug,
       volume = {469},
       number = {4},
        pages = {3994-4011},
          doi = {10.1093/mnras/stx1103},
archivePrefix = {arXiv},
       eprint = {1705.03305},
 primaryClass = {astro-ph.EP},
       adsurl = {https://ui.adsabs.harvard.edu/abs/2017MNRAS.469.3994B}
}

@ARTICLE{Schneider2021,
       author = {{Schneider}, Aaron David and {Bitsch}, Bertram},
        title = "{How drifting and evaporating pebbles shape giant planets. I. Heavy element content and atmospheric C/O}",
      journal = {\aap},
         year = 2021,
        month = oct,
       volume = {654},
          eid = {A71},
        pages = {A71},
          doi = {10.1051/0004-6361/202039640},
archivePrefix = {arXiv},
       eprint = {2105.13267},
 primaryClass = {astro-ph.EP},
       adsurl = {https://ui.adsabs.harvard.edu/abs/2021A&A...654A..71S}
}

@ARTICLE{Mah2023,
       author = {{Mah}, Jingyi and {Bitsch}, Bertram and {Pascucci}, Ilaria and {Henning}, Thomas},
        title = "{Close-in ice lines and the super-stellar C/O ratio in discs around very low-mass stars}",
      journal = {\aap},
         year = 2023,
        month = sep,
       volume = {677},
          eid = {L7},
        pages = {L7},
          doi = {10.1051/0004-6361/202347169},
archivePrefix = {arXiv},
       eprint = {2308.15128},
 primaryClass = {astro-ph.EP},
       adsurl = {https://ui.adsabs.harvard.edu/abs/2023A&A...677L...7M}
}

@ARTICLE{Perotti2023,
       author = {{Perotti}, G. and {Christiaens}, V. and {Henning}, Th. and {Tabone}, B. and {Waters}, L.~B.~F.~M. and {Kamp}, I. and {Olofsson}, G. and {Grant}, S.~L. and {Gasman}, D. and {Bouwman}, J. and {Samland}, M. and {Franceschi}, R. and {van Dishoeck}, E.~F. and {Schwarz}, K. and {G{\"u}del}, M. and {Lagage}, P.-O. and {Ray}, T.~P. and {Vandenbussche}, B. and {Abergel}, A. and {Absil}, O. and {Arabhavi}, A.~M. and {Argyriou}, I. and {Barrado}, D. and {Boccaletti}, A. and {Caratti o Garatti}, A. and {Geers}, V. and {Glauser}, A.~M. and {Justannont}, K. and {Lahuis}, F. and {Mueller}, M. and {Nehm{\'e}}, C. and {Pantin}, E. and {Scheithauer}, S. and {Waelkens}, C. and {Guadarrama}, R. and {Jang}, H. and {Kanwar}, J. and {Morales-Calder{\'o}n}, M. and {Pawellek}, N. and {Rodgers-Lee}, D. and {Schreiber}, J. and {Colina}, L. and {Greve}, T.~R. and {{\"O}stlin}, G. and {Wright}, G.},
        title = "{Water in the terrestrial planet-forming zone of the PDS 70 disk}",
      journal = {\nat},
         year = 2023,
        month = aug,
       volume = {620},
       number = {7974},
        pages = {516-520},
          doi = {10.1038/s41586-023-06317-9},
archivePrefix = {arXiv},
       eprint = {2307.12040},
 primaryClass = {astro-ph.EP},
       adsurl = {https://ui.adsabs.harvard.edu/abs/2023Natur.620..516P}
}

@ARTICLE{Banzatti2023,
       author = {{Banzatti}, Andrea and {Pontoppidan}, Klaus M. and {Carr}, John S. and {Jellison}, Evan and {Pascucci}, Ilaria and {Najita}, Joan R. and {Romero-Mirza}, Carlos E. and {{\"O}berg}, Karin I. and {Kalyaan}, Anusha and {Pinilla}, Paola and {Krijt}, Sebastiaan and {Long}, Feng and {Lambrechts}, Michiel and {Rosotti}, Giovanni and {Herczeg}, Gregory J. and {Salyk}, Colette and {Zhang}, Ke and {Bergin}, Edwin A. and {Ballering}, Nicholas P. and {Meyer}, Michael R. and {Bruderer}, Simon and {Jdiscs Collaboration}},
        title = "{JWST Reveals Excess Cool Water near the Snow Line in Compact Disks, Consistent with Pebble Drift}",
      journal = {\apjl},
         year = 2023,
        month = nov,
       volume = {957},
       number = {2},
          eid = {L22},
        pages = {L22},
          doi = {10.3847/2041-8213/acf5ec},
archivePrefix = {arXiv},
       eprint = {2307.03846},
 primaryClass = {astro-ph.EP},
       adsurl = {https://ui.adsabs.harvard.edu/abs/2023ApJ...957L..22B}
}

@ARTICLE{Paardekooper2006,
       author = {{Paardekooper}, S.-J. and {Mellema}, G.},
        title = "{Dust flow in gas disks in the presence of embedded planets}",
      journal = {\aap},
         year = 2006,
        month = jul,
       volume = {453},
       number = {3},
        pages = {1129-1140},
          doi = {10.1051/0004-6361:20054449},
archivePrefix = {arXiv},
       eprint = {astro-ph/0603132},
 primaryClass = {astro-ph},
       adsurl = {https://ui.adsabs.harvard.edu/abs/2006A&A...453.1129P}
}

@ARTICLE{Lambrechts2014,
       author = {{Lambrechts}, M. and {Johansen}, A. and {Morbidelli}, A.},
        title = "{Separating gas-giant and ice-giant planets by halting pebble accretion}",
      journal = {\aap},
         year = 2014,
        month = dec,
       volume = {572},
          eid = {A35},
        pages = {A35},
          doi = {10.1051/0004-6361/201423814},
archivePrefix = {arXiv},
       eprint = {1408.6087},
 primaryClass = {astro-ph.EP},
       adsurl = {https://ui.adsabs.harvard.edu/abs/2014A&A...572A..35L}
}

@ARTICLE{Bitsch2018,
       author = {{Bitsch}, Bertram and {Morbidelli}, Alessandro and {Johansen}, Anders and {Lega}, Elena and {Lambrechts}, Michiel and {Crida}, Aur{\'e}lien},
        title = "{Pebble-isolation mass: Scaling law and implications for the formation of super-Earths and gas giants}",
      journal = {\aap},
         year = 2018,
        month = apr,
       volume = {612},
          eid = {A30},
        pages = {A30},
          doi = {10.1051/0004-6361/201731931},
archivePrefix = {arXiv},
       eprint = {1801.02341},
 primaryClass = {astro-ph.EP},
       adsurl = {https://ui.adsabs.harvard.edu/abs/2018A&A...612A..30B}
}

@misc{pandas,
author = {McKinney, Wes},
year = {2010},
month = {01},
title = {Data Structures for Statistical Computing in Python},
doi = {10.25080/Majora-92bf1922-00a}
}
\bibliographystyle{aasjournalv7}

\end{document}